\documentclass[usenatbib]{mnras}
\usepackage{graphicx}
\usepackage[fleqn]{amsmath}
\usepackage[T1]{fontenc}
\usepackage{ae,aecompl}
\usepackage{amssymb,amsfonts,hyperref,color,subfig}
\hypersetup{linkcolor=red,citecolor=blue,filecolor=blue,urlcolor=magenta}
\usepackage{etoolbox}
\makeatletter
\patchcmd\@combinedblfloats{\box\@outputbox}{\unvbox\@outputbox}{}{%
   \errmessage{\noexpand\@combinedblfloats could not be patched}%
}%
 \makeatother
\usepackage[dvipsnames]{xcolor}

\title[Multiple dust rings with a migrating planet]{Intermittent planet migration and the formation of multiple dust rings and gaps in protoplanetary disks}

\author[Wafflard-Fernandez \& Baruteau]{Gaylor Wafflard-Fernandez,$^{1}$\thanks{E-mail: gwafflard@irap.omp.eu}
Cl{\'e}ment Baruteau$^{1}$
\\
$^{1}$IRAP, Universit{\'e} de Toulouse, CNRS, UPS, F-31400 Toulouse, France\\
}

\date{Accepted 2020 February 6. Received 2020 February 6; in original form 2019 August 13}

\pubyear{T2020}
\begin{document}
\label{firstpage}
\pagerange{\pageref{firstpage}--\pageref{lastpage}}
\maketitle

\begin{abstract}
A key challenge for protoplanetary disks and planet formation models is to be able to make a reliable connection between observed structures in the disks emission, like bright and dark rings or asymmetries, and the supposed existence of planets triggering these structures. The observation of N dark rings of emission is often interpreted as evidence for the presence of N planets which clear dust gaps around their orbit and form dust-trapping pressure maxima in the disk. The vast majority of the models that studied the impact of planets on the dynamics of dust and gas in a protoplanetary disk assumed planets on fixed orbits. Here we go a different route and examine how the large-scale inward migration of a single planet structures the dust content of a massive disk. In many circumstances, the migration of a partial gap-opening planet with a mass comparable to Saturn is found to run away intermittently. By means of 2D gas and dust hydrodynamical simulations, we show that intermittent runaway migration can form multiple dust rings and gaps across the disk. Each time migration slows down, a pressure maximum forms beyond the planet gap that traps the large dust. Post-processing of our simulations results with 3D dust radiative transfer calculations confirms that intermittent runaway migration can lead to the formation of multiple sets of bright and dark rings of continuum emission in the (sub)millimeter beyond the planet location.
\end{abstract}


\section{Introduction}
\label{sec:intro}
Recent observations of substructures in protoplanetary disks have challenged the classical picture of disks being smooth continuous objects. A fair fraction of spatially resolved disks show indeed substructures, including spirals, axisymmetric and asymmetric rings when observed from the near-IR to the (sub-)mm. In particular, sequences of dark rings (gaps) and bright rings are the most frequent substructures revealed by radio interferometry with ALMA. In \citet{Long2018a}, 40$\%$ (12/32) of their sample feature axisymmetric rings and gaps. For some disks, such structures have also been observed in SPHERE near-IR polarized observations \citep[e.g.,][]{Avenhaus2018}.

If we focus on Disks with Annular Substructures (DAS) in the (sub-)mm continuum from the Disk Substructures at High Angular Resolution Project (DSHARP) in addition to HL Tau, TW Hya and other DAS listed in Table 5 of \citet{Huang2018}, we see that around 30$\%$ (12/42) of DAS are single-ring systems and 70$\%$ (30/42) are multiple-ring systems. This classification is based on the current resolution at which disks are observed. A better resolution could possibly reveal more rings. This is the case for example of the disk around HD 169142, for which the bright ring observed at $\sim0\farcs6$ with a $\sim0\farcs2$ beam by \citet{Fedele2017} turns out to be composed of three narrow rings when observed with an angular resolution ten times better \citep{Perez2019}. Another bias is the choice of the targets: in the DSHARP sample, targets were deliberately composed of massive and extended disks, and \citet{Long2018a} suggest that multiple-ring disks are more massive and larger. This could explain the seemingly large current fraction of multiple-rings disks. In any case, the mechanisms behind the presence of annular substructures should be sufficiently general to occur in a significant fraction of protoplanetary disks and should be able to generate multiple rings.

Bright and dark rings observed in the (sub-)mm are often interpreted as dust over-densities and under-densities. However, generally, radial variations of the observed intensity are not necessarily related to radial variations of the dust density. When scattering is negligible, the specific intensity at frequency $\nu$ ($I_\nu$) is linked to the dust temperature $T_{\rm d}$, the dust's surface density $\Sigma_{\rm d}$, and the dust's absorption opacity $\kappa_\nu$ \citep[see, e.g., the review by][his equation 2]{Andrews15}. Thus, radial variations of $I_\nu$ can be due to radial variations of $T_{\rm d}$, $\Sigma_{\rm d}$ and/or $\kappa_\nu$. Any mechanism able to induce radial variations of these quantities could in principle explain bright and dark rings of emission. Still, we will concentrate in the following on mechanisms that cause radial variations of $\Sigma_{\rm d}$. One such mechanism is dust pile-up due to sintering-induced fragmentation \citep{Okuzumi2016}. Sintering, which bonds dust particles together, occurs efficiently for micro-metric dust monomers slightly outside the snow lines of volatile species. When sintered aggregates collide, they are prone to bouncing, fragmentation and erosion because the collision energy cannot be dissipated, which decreases their sticking efficiency. Sintering thus decreases the size of dust aggregates and therefore their drift velocity, thereby causing a dust pile-up slightly beyond the snow lines.

Several disks show very thin bright rings intercalated between dark rings in the continuum emission, such as HD 163296 or AS 209 \citep{Huang2018}. Assuming these bright rings are dust over-densities, their thinness suggests that they are located at positions of equilibrium for the dust, where radial drift is stalled. These so-called dust traps can be naturally obtained at pressure maxima in a disk. Before mentioning several ways to form pressure bumps in a disk, we note that a pressure maximum is susceptible to the Rossby Wave Instability \citep{Lovelace1999}, which may trigger the formation of one or several vortices that also trap dust in the azimuthal direction \citep{Li2001, BargeSommeria1995}. Dust trapping in vortices is often invoked to explain the asymmetric crescent-shaped structures in the disks continuum emission at radio wavelengths \citep[e.g.,][]{Baruteau2019}.

Various mechanisms have been proposed to generate pressure bumps in protoplanetary disks. One is via the radial changes in the accretion efficiency at the transition between magnetically active regions in disks (where the magneto-rotational instability, or MRI, sets in) and magnetically dead zones (where the MRI is quenched due to non-ideal magneto-hydrodynamic effects; \citealp[see, e.g.,][]{Varniere2006, Flock15, Lyra15}). Moreover, when the disk is threaded by a weak magnetic field, a linear instability driven by magneto-hydrodynamic winds can induce zonal flows which take the form of axisymmetric rings with alternating minima and maxima in the magnetic pressure, which are associated with maxima and minima in the thermal pressure \citep{Bethune2017, RiolsLesur19}. Another popular scenario is planets. A planet that is sufficiently massive to carve an annular gap around its orbit will form pressure maxima at both edges of its gap due to the deposition of the angular momentum flux carried away by the wakes of the planet \citep{BaruteauPP6, Bae2016}.

Models that invoke planets to explain annular substructures in disks observations almost always assume a planet on a fixed circular orbit at the middle of each dark ring \citep[see, e.g.,][for the HL Tau disk]{Dipierro2015}. Put another way, N gaps would correspond to N planets in a disk (hereafter, NGNP model). Some models have actually proposed to use the radial width of a dark ring to estimate the mass of the planet which created that ring \citep[e.g.,][]{Dipierro17,Zhang2018}. The presence of planets in dark rings still eludes direct detection, with the notable exception of the planet companion in the large gap of the PDS 70 transition disk \citep{Keppler2018}. Other studies have also explored the idea that, in low-viscosity disks, a planet on a fixed circular orbit may form a secondary gap inside its orbit, and even a tertiary gap in some cases \citep{Bae2017multiple, Dong2017multiple, Zhang2018}.

In this work we examine how a single planet that migrates in its disk impacts the dust's spatial distribution and continuum emission. A planet will impact the dust structures differently depending on how the planet's migration rate compares with the dust's drift rate. A few studies have recently assessed the effect of planetary migration on the dust dynamics, in particular the presence and location of dust rings on both sides of the planet, the dependence on the dust's size distribution, and how the position of a dust ring relative to the planet can be an indicator of the planet's migration speed \citep[e.g.,][]{Meru2019,Nazari2019,Perez2019,Weber2019}. The parameter space is wide, however, as it involves a range of planet masses, disk's physical properties (density, temperature profiles etc.) and dust sizes. We shall focus on the migration of moderately massive planets (typically around a Saturn mass) in fairly massive disks, which are subject to rapid runaway migration \citep{Masset2003}. Such planets can undergo multiple stages of intermittent runaway migration \citep{Lin2010,McNally2019}. We find that these multiple stages of intermittent runaway migration lead to the formation of several dust rings and therefore sequences of dark and bright rings in the (sub-)mm continuum.

The plan of this paper is the following. In Section~\ref{sec:methods}, we describe the physical model and numerical methods of the hydrodynamical simulations and the radiative transfer calculations. The results are then presented in Section~\ref{sec:results}, where we focus on two fiducial cases. A discussion and a summary follow in Section~\ref{sec:conclusion}.

\section{Physical model and numerical methods}
\label{sec:methods}

\subsection{Hydrodynamical Simulations}
We performed 2D hydrodynamical simulations of the gas and dust of a protoplanetary disk with an embedded planet. Simulations were carried out with the code Dusty FARGO-ADSG, which is an extension of the grid-based code \href{http://fargo.in2p3.fr/-FARGO-ADSG-}{FARGO-ADSG} \citep{Masset2000,BaruteauMasset2008a,BaruteauMasset2008b} with dust grains modeled as Lagrangian test particles \citep{Baruteau2016,Fuente2017}. We adopt a polar coordinate system ($r$, $\theta$), with $r$ the radial cylindrical coordinate measured from the central star and $\theta$ the azimuthal angle. We detail below the physical model and numerical setup for the gas, dust and planet in our simulations. A short summary of the main parameters of the simulations is given in Table~\ref{table:parameters}.
\begin{table}
\centering
\caption{\label{table:parameters} Main parameters of the hydrodynamical simulations.}
\begin{tabular}{lr}
\hline
\hline
Parameter                       & Value(s)           \\
\hline
Code's unit of length                      & $r_0$ = 10 au \\
Disk's aspect ratio ($h$) at $r_0$           & \{0.05,0.06\}         \\
Toomre-Q parameter at $r_0$         & $\in [5.3 - 64]$   \\
Alpha turbulent viscosity                & \{$10^{-4}$,$10^{-3}$\}         \\
Planet's initial location                   & $r_{\rm p,0} = 2r_0 = 20$ au         \\
Planet-to-primary mass ratio         & $\sim$1.46 $h^3(r_{\rm p,0})$  \\
Dust's size range & $\in [10\,\mu{\rm m} - 10\,{\rm cm}]$\\
Dust's internal density & 2 g cm$^{-3}$\\
\hline
\end{tabular}
\end{table}

\subsubsection{Gas}
\label{sec:gas}
We solve the continuity and the Navier-Stokes equations for the gas on a polar mesh centered on the star, taking into account the indirect terms due to the acceleration of the star by the planet and the disk. In the radial direction, the grid stretches from 2 au to 40 au.

We assume for simplicity a locally isothermal equation of state for the gas. This means that the gas temperature, or equivalently the sound speed, is fixed in time but varies with $r$. This assumption is generally more appropriate to the outer parts of protoplanetary disks, where the radiative cooling and/or diffusion timescales are short compared to the orbital timescale. However, we will see in Section~\ref{sec:impact_nrj} that allowing the gas temperature to vary with time by solving the energy equation does not qualitatively change our results. In our simulations with a locally isothermal equation of state, the aspect ratio $h(r)$, which is the ratio between the sound speed and the Keplerian speed, is chosen as $h(r) = h_0 \times \left(r/r_0\right)^{0.15}$, with $r_0=10~\textrm{au}$ the code's units of length. Two values for the aspect ratio at $r_0$ are adopted : $h_0 = 0.05$ and $0.06$, which respectively correspond to a temperature of 63 K and 91 K for a Sun mass star and a solar composition (mean molecular weight $\mu_{\sun}=2.4$).

We model the radial turbulent transport of the disk's angular momentum by an equivalent kinematic viscosity. We parameterize this viscosity as $\nu=\alpha c_{\rm S} H$, with $\alpha$ a dimensionless constant, $c_{\rm S}$ the sound speed, and $H = h(r) \times r$ the disk pressure's scale height \citep{Shakura1973}. The alpha turbulent viscosity at a few to a few tens of au in the disks midplane is highly uncertain. Non-ideal magneto-hydrodynamic simulations show that it may vary from a few $\times 10^{-5}$ to as high as a few percent, depending on the amplitude and sign of the vertical magnetic field that threads the disk \citep{Simon15,Bethune2017}. Modeling of the ring-like structures in the (sub)millimeter continuum emission of some protoplanetary disks suggests that the midplane alpha at few tens of au should be $\lesssim 10^{-4}$ in order to reproduce the rings sharpness \citep[e.g.,][]{Pinte2016,Perez2019}. These rather low values are overall consistent with observations of the non-thermal broadening in protoplanetary disks \citep{Flaherty2015}. To reflect this uncertainty, we will adopt two values of $\alpha$ in this study: $10^{-4}$ and $10^{-3}$. We will see that, independently of its effect on planetary migration, the value of $\alpha$ has a clear impact on the dust structures that form because of disk-planet interactions, such as their lifetime, their radial width, or their degree of axisymmetry.

The initial gas surface density profile is $\Sigma_0(r) = \Sigma_0 \times \left(r/r_0\right)^{-1}$, with $\Sigma_0$ the gas surface density at $r_0$. We adopt in this work three values for $\Sigma_0$: $3 \times 10^{-4}$, $10^{-3}$ and $3 \times 10^{-3}$ in code units, which, by choosing a code's unit of mass of 1 $M_{\sun}$, correspond to about 27, 89 and 267 $\textrm{g\;cm}^{-2}$ at $r_0 = 10$ au, respectively. For these three values of $\Sigma_0$, the disk-to-star mass ratio amounts to $0.007$, $0.02$ or $0.07$, respectively, while the Toomre Q-parameter at $r_0$, which is proportional to $h_0$, initially equals $53$, $16$ and $5.3$, respectively, for $h_0 = 0.05$. Gas self-gravity is included in our simulations, and the disk gas is stable against the gravitational instability throughout the computational domain. Because we perform 2D simulations, a softening length of $0.3H(r)$ is used in the calculation of the self-gravitating acceleration to mimic the vertical extent of the disk \citep{BaruteauMasset2008b,Muller2012}.

In the azimuthal direction, the grid extends from 0 to 2$\pi$ and is paved by 900 cells that are evenly spaced. In the radial direction, we use 600 radial cells with a logarithmic spacing, which is necessary for the self-gravitating acceleration to read as a convolution product and be computed via Fast Fourier Transforms \citep{BaruteauMasset2008b}. For the boundary conditions, we use wave-killing zones near the inner and outer edges of the grid to avoid reflections of the planet wakes. These zones extend from 2 to 3 au, and from 35 to 40 au. The surface density and radial velocity of the gas are damped towards the radial profiles obtained by calculating the pure viscous evolution of the disk on a 1D grid simultaneously to the gas equations solved on the polar grid (the azimuthal velocity of the gas is damped towards its axisymmetric instantaneous profile). This is basically the same boundary condition as that implemented in FARGO-3D by \citet{Benitez-Llambay2016}. It is a very simplified way to model the viscous evolution of a global disk in the absence of planets. Given the viscosity adopted in our disk models ($\alpha \leq 10^{-3}$) and the duration of our runs (a few thousand planet orbits, at most), this boundary condition is, in practice, nearly identical to damping the gas density and radial velocity towards their initial radial profile. Our 1D grid has 1000 cells logarithmically spaced between 0.4 and 300 au, and the gas surface density is set to zero at the edges of the 1D grid. The choice of boundary conditions does not change significantly the results as long as the planet remains further than $\sim 4.5$ au.

\subsubsection{Dust}
\label{sec:methods_dust}
Dust is modeled as Lagrangian super-particles which feel the gravity of the star, of the disk gas (since gas self-gravity is included), of the planet, and gas drag. However, because we do not take into account dust self-gravity, collisions, dust growth nor fragmentation, particles do not feel each other. Moreover there is no dust back-reaction onto the gas. We will discuss this assumption in the simulations results in Sections~\ref{sec:case1} and~\ref{sec:case2}. To model the effects of gas turbulence, stochastic kicks are applied to the position ($r_{\rm d}$, $\theta_{\rm d}$) of the dust particles at each hydrodynamical timestep of the simulation (kicks depend on the local kinematic viscosity of the gas and Stokes number of the dust, see \citealp{Ataiee18} for more details).

Particles are assumed to have an internal density of $2~\textrm{g\;cm}^{-3}$, independently of their size. They have a size distribution such that $n(s)ds$, which corresponds to the number of super-particles in the size interval $[s,s+ds]$, is a power-law function of the particle size $s$, going from $s_{\rm min} = 10~\mu$m to $s_{\rm max} = 10$ cm. The power-law exponent is set to -1 in order to have approximately the same number of particles per decade of size. Even though we need a (more) realistic size distribution exponent to compute the dust emission through radiative transfer calculations, the choice of the power-law exponent has no impact on the hydrodynamical simulations since dust feedback on the gas is discarded. The maximum particles size ($s_{\rm max}$) is consistent with the results of 1D models of dust growth, drift and fragmentation in planet-less disks having similar physical properties as the background disk in our simulations \citep{Birnstiel2012}. The number of particles and their initial location will be specified in Sections~\ref{sec:case1} and~\ref{sec:case2}.

\subsubsection{Planet ($\mathcal{P}$)}
The planet, which we will refer to as $\mathcal{P}$ in this paper, is set initially at $r_{\rm p,0}$ = 2$r_0$ = 20 au from the star. This work focuses on how the migration of $\mathcal{P}$ due to its interactions with the disk gas impacts the formation of dust rings by the spiral wakes of $\mathcal{P}$. The mass of $\mathcal{P}$ is therefore chosen such that its wakes can form pressure maxima on either side of the planet's orbit, where dust particles can be trapped. We have chosen to study planets that carve a partial gap in the gas around their orbit, and which can be subject to fast runaway type III migration if the disk is sufficiently massive \citep{Masset2003}. Accretion on the planet is discarded in this study.

The range of planet-to-primary mass ratios ($q_{\rm p}$) for which partial gap opening occurs depends on the disk's aspect ratio and turbulent viscosity near the planet's location. Throughout this work, unless otherwise stated, results are obtained for $q_{\rm p} \approx 1.46 h^3(r_{\rm p,0})$. More specifically, the simulations with $h_0 = 0.05$ have $q_{\rm p} = 2.5\times10^{-4}$, those with $h_0 = 0.06$ have $q_{\rm p} = 4.3\times10^{-4}$. These correspond to planets of $\sim$0.9 and $\sim$1.5 the mass of Saturn orbiting Solar-mass stars. The mass of the planet is gradually increased over its first 5 orbits, yet it is allowed to migrate from the beginning of the simulations. A softening length of $0.6H(r_{\rm p})$ is used to calculate the planet acceleration on the gas, where $r_{\rm p}$ denotes the (time varying) orbital radius of the planet.

\subsection{Dust radiative transfer calculations}
\label{sec:RTsetup}
We post-process our results of simulations with dust radiative transfer calculations to infer the continuum emission that results from the dust's annular substructures obtained in our simulations. The radiative transfer calculations are carried out with the code \href{http://www.ita.uni-heidelberg.de/~dullemond/software/radmc-3d}{RADMC3D} (version 0.41, \citealp{Dullemond2015}). Synthetic maps of continuum emission are computed for a face-on disk at $\lambda=1.3$ mm, which is the same wavelength as in the ALMA survey of disks undertaken by DSHARP \citep{Huang2018}. The procedure to compute RADMC3D synthetic images of the continuum emission from the outputs of the code Dusty FARGO-ADSG is described in \citet{Baruteau2019} and makes use of the publicly available python program \href{https://github.com/charango/fargo2radmc3d}{fargo2radmc3d}. We recap here the main computing steps for convenience.

The dust's spatial distribution obtained in the hydrodynamical simulations is the key quantity to compute the continuum emission. We first derive the surface density of dust in 30 size bins logarithmically spaced between $s_{\rm min}$ and $s_{\rm max}$, the minimum and maximum particle sizes in the hydrodynamical simulations. This is done by assuming a size distribution $n(s) \propto s^{-3.5}$ and a total dust mass $M_{\rm dust} = 10^{-2}M_{\rm gas}$, where $M_{\rm gas}$ is the mass of the disk gas in the simulations. We then obtain the dust's volume density in each size bin and in each grid cell of the 3D spherical grid used by RADMC3D, which is the 2D polar grid of the simulations extended in colatitude with 36 cells spanning $\pm 2H$ around the midplane (a logarithmic spacing is used in colatitude with finer grid cells near the midplane). For this, we assume that the vertical distribution of the dust's volume density is Gaussian for each size bin, with a characteristic scale height equal to $H \times \left( 1 + \rm{St}_{\rm bin} / \alpha \right)^{-1/2}$, where $\rm{St}_{\rm bin}$ is the average Stokes number of dust particles in each size bin.

Opacities are calculated using Mie theory, assuming a dust mixture comprised of $60\%$ astrosilicates and $40\%$ water ices, which corresponds to a mean internal density of $\sim$1.7 g cm$^{-3}$, close to that assumed in the hydrodynamical simulations. We use the Bruggeman formula to obtain the optical constants of the mixture (the optical constants of water ices are from the \href{http://www.astro.uni-jena.de/Laboratory/Database/databases.html}{Jena database}, those of astrosilicates are from \citealp{Draine1984}).

Armed with the dust's volume density and opacities, the radiative transfer calculations proceed in two steps: first, the dust temperature is computed via a thermal Monte-Carlo calculation, next the specific intensity of continuum emission is calculated by solving the equation of radiative transfer via photon ray-tracing. We assume the disk to be located at 150 pc, with zero inclination relative to the line of sight, and a central star with a radius of 2 R$_{\odot}$ and effective temperature of 7000 K. We use $10^9$ photon packages for the thermal Monte-Carlo and ray-tracing calculations. Scattering off dust grains is discarded as it is found to have a small impact on our final synthetic images. We have checked this by dedicated RADMC3D calculations including anisotropic scattering with the Henyey-Greenstein approximation for the scattering phase function. For the simulation presented in Section~\ref{sec:case2}, inclusion of anisotropic scattering results in $\sim$15\% relative difference for the flux of continuum emission compared to a calculation without scattering. Since our disk model is rather optically thick at 1.3 mm, discarding scattering significantly decreases the computation time of the radiative transfer calculations.

Finally, knowing the disk distance, the specific intensity is convolved with a $0.02"$ beam in order to obtain a synthetic flux map of continuum emission. In some of the synthetic maps presented in Section~\ref{sec:results}, white noise is included by adding at each pixel of the synthetic map of specific intensity a random number following a Gaussian distribution with zero mean and a standard deviation set to 50 $\mu$Jy/beam. This standard deviation is similar to the rms noise level obtained in recent ALMA observations at 1.3 mm and at similar resolution \citep[see, e.g.,][]{Perez2019}.

\section{Results}
\label{sec:results}
We begin this section with an overview of our results of hydrodynamical simulations in Section~\ref{sec:overview}, which delineates the region of parameter space for which intermittent planetary migration is obtained. We then focus in Sections~\ref{sec:case1} and~\ref{sec:case2} on two simulations in which intermittent migration results in the formation of multiple dark and bright rings of emission at millimeter wavelengths.

\subsection{Overview}
\label{sec:overview}

\begin{figure}
\centering
\resizebox{\hsize}{!}
{
\includegraphics[width=0.99\hsize]{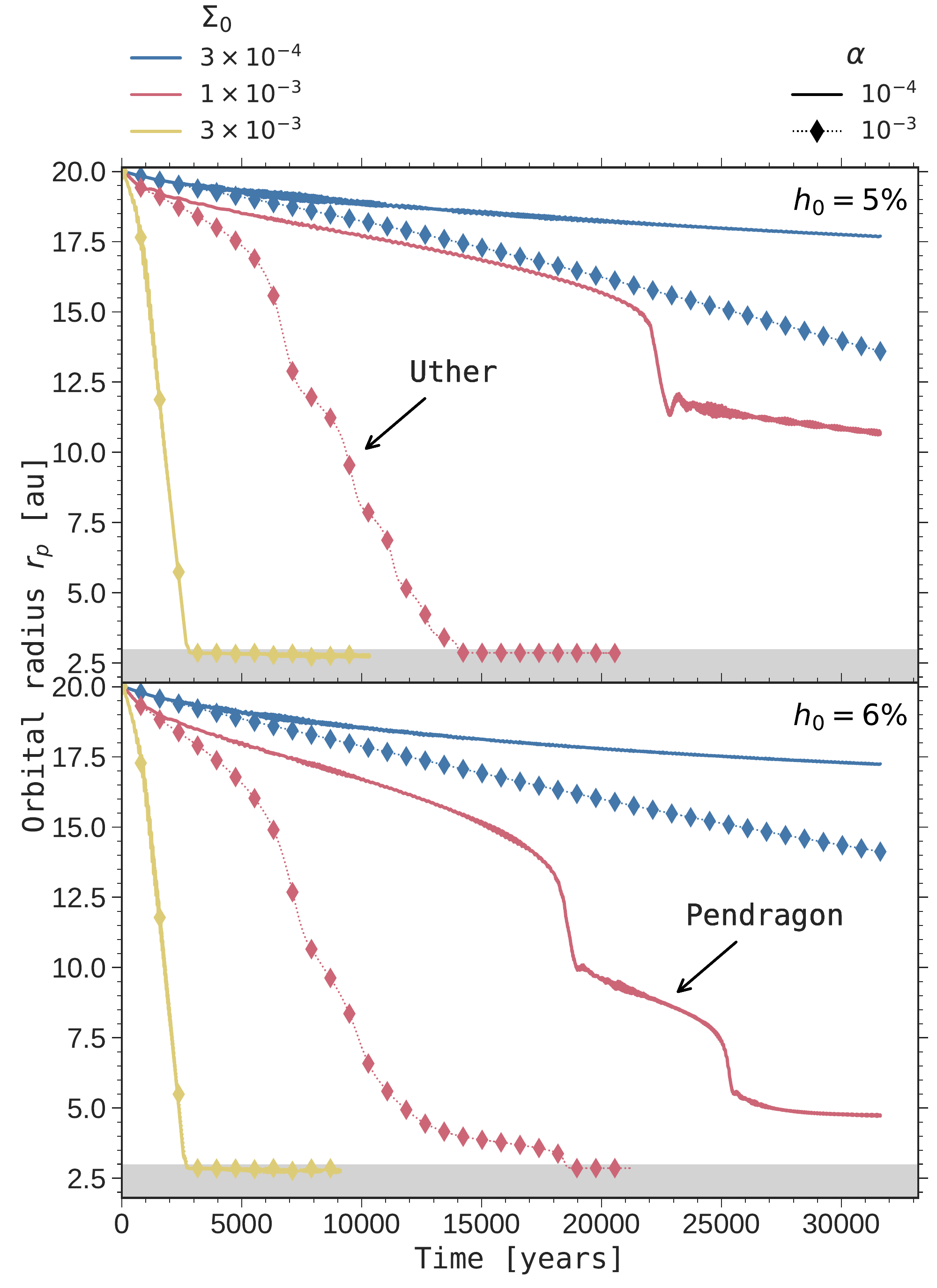}
}
\caption{Overview of our results of hydrodynamical simulations: time evolution of the planet's orbital radius for three values of the gas surface density ($\Sigma_0$), two values of the alpha turbulent viscosity ($\alpha$) and of the disk's aspect ratio ($h_0$). \textbf{Top:} $h_0=5\%$. \textbf{Bottom:} $h_0=6\%$. The gray bands correspond to the inner wave-killing zone.}
\label{fig:general}
\end{figure}

We present in Fig.~\ref{fig:general} a general view of twelve simulations characterized by the parameters $h_0$, $\alpha$ and $\Sigma_0$ described in Section~\ref{sec:gas}. In this figure, we plot the evolution of the planet's orbital radius as a function of time, with a panel for each value of $h_0$. For both values of $h_0$, we remark that the graphs can be separated into three specific areas depending on $\Sigma_0$:

\begin{itemize}
\item \textit{Fast migration:} for the most massive disk model ($\Sigma_0 = 3 \times 10^{-3}$), the four curves (in yellow) almost lie on top of one another, which shows that the migration rate hardly depends on $\alpha$ and $h_0$. It indicates that migration is mostly driven by a strong dynamical corotation torque, rather than by the the wave (Lindblad) torque and the static corotation torque (both of which scale as $h^{-2}_0$ for low-mass planets). The very fast inward migration stalls when the planet ($\mathcal{P}$) reaches the inner damping zone of the grid, where its migration is no longer determined by the physical properties of the disk. This migration regime is so fast that the interaction between $\mathcal{P}$ and the dust particles remains limited, as is illustrated in the first row of Fig.~\ref{fig:appendix} in Appendix~\ref{sec:appendix_a}. As $\mathcal{P}$ moves inwards, most dust particles inside the orbit of $\mathcal{P}$ perform indeed an outer horseshoe U-turn relative to $\mathcal{P}$ and end up forming a dust band behind $\mathcal{P}$ that is similar to the initial one. Some of the particles embarking onto horseshoe U-turns also become trapped around the L4 Lagrange point in front of $\mathcal{P}$ in the azimuthal direction. Once the dust band is formed again, it drifts and diffuses radially due to gas drag and turbulence, respectively.
\medskip
\item \textit{Slow migration:} for the least massive disk model ($\Sigma_0 = 3 \times 10^{-4}$, blue curves), migration is slow and the dust rings generated by $\mathcal{P}$ display little evolution over the duration of the simulations. $\mathcal{P}$ actually forms three dust rings: one on each side of its partial gap in the gas, and one that stays coorbital with $\mathcal{P}$. As $\mathcal{P}$ moves inwards, the inner and coorbital dust rings move inward with $\mathcal{P}$ while the outer ring roughly stays at its initial location \citep[][see also Section~\ref{sec:dsd_2} and the second row of Fig.~\ref{fig:appendix} in Appendix~\ref{sec:appendix_a}]{Perez2019,Meru2019}. This implies that the mutual separation between the three rings increases as $\mathcal{P}$ migrates. This effect may well be observed in the outer parts of the HD 169142 disk, where the radial asymmetry between the three fine bright rings located between $\sim0\farcs5$ and $\sim0\farcs7$ has been interpreted as the signature of an inward-migrating mini-Neptune planet \citep{Perez2019}.
\medskip
\item \textit{Intermittent migration:} the intermediate curves for $\Sigma_0 = 10^{-3}$ (in red) show non-steady (or non-smooth) orbital evolution, with several short stages of runaway migration which impart a stair-case shape to the curves. \citet{Lin2010} and \citet{McNally2019} have reported results of hydrodynamical simulations with similar multiple episodes of runaway migration, which were obtained in inviscid or low-viscosity disk models. These intermittent stages of runaway migration are related to the time evolution of the coorbital vorticity-weighted mass deficit of the planet \citep{Masset2003}. This quantity basically compares the inverse vortensity\footnote{The gas vortensity, or potential vorticity, refers in this study to the ratio between the vorticity and the surface density, with the vorticity being the z-component of the curl of the 2D velocity.} of the gas crossing the planet's orbit with that of the gas trapped in the planet's horseshoe region (see Eq.~\ref{eq:deficit} below). Although it has the dimension of a mass, the coorbital vorticity-weighted mass deficit should rather be considered a coorbital vortensity deficit, especially for the low-mass planets considered in our study \citep{Paardekooper2014}. Note in particular that $\delta m$ may be either positive or negative, depending on the background vortensity profile. We will now call $\delta m$ the coorbital vortensity deficit. When $\delta m$ exceeds the mass of the planet ($M_{\rm p}$) plus that of its circumplanetary disk ($M_{\rm cpd}$), $\mathcal{P}$ enters a regime of runaway migration\footnote{We have checked that $M_{\rm cpd}$ is always smaller than $M_{\rm p}$ in our simulations, by typically an order of magnitude. Thus, we will now simply use $\delta m > M_{\rm p}$ to define the onset of runaway migration.} \citep{Masset2003,Masset08}. However, as will be shown via one of our simulations in Section~\ref{sec:case1}, $\delta m$ may decrease during runaway migration, with the consequence that runaway ceases and the migration of $\mathcal{P}$ proceeds at a slower pace. We will show that during this phase of reduced migration, the vortensity inside the horseshoe region gets mixed and runaway can start again.
\end{itemize}

In the following, we focus on the results of two simulations, which are labeled Uther and Pendragon in Fig.~\ref{fig:general}, for which $\mathcal{P}$ displays intermittent runaway migration. The results of both simulations are described in Sections~\ref{sec:case1} and~\ref{sec:case2}.

\subsection{Uther simulation ($h_0 = 0.05$, $\alpha = 10^{-3}$)}
\label{sec:case1}
The simulation that we refer to as Uther displays four episodes of runaway migration in the upper panel of Figure~\ref{fig:general}. We begin in Section~\ref{sec:cvd} by showing that these intermittent stages of runaway migration are due to oscillations in the planet's coorbital vortensity deficit. We then examine the impact of these successive jolts of the planet on the dust's spatial distribution in Section~\ref{sec:dsd_1}, and on the dust's thermal emission at millimeter wavelengths in Section~\ref{sec:dte_1}. The main parameters of the simulation are summarized in Table~\ref{table:case1_parameters}.

\begin{table}
\centering
\caption{\label{table:case1_parameters}Parameters of simulation Uther (Section~\ref{sec:case1})}
\begin{tabular}{lr}
\hline
\hline
Parameter                           & Value           \\
\hline
Planet-to-primary mass ratio $q_{\rm p}$                  & $2.5\times10^{-4}$         \\
Disk's aspect ratio $h_0$ at 10 au                       & 0.05       \\
Alpha turbulent viscosity $\alpha$        & $10^{-3}$   \\
Gas surface density $\Sigma_0$ at 10 au [code units]        & $10^{-3}$   \\
\hline
Number of dust super-particles  & 50000\\
Dust's initial location  & $\in [14-21]$ au\\
\hline
\end{tabular}
\end{table}

\subsubsection{Coorbital vortensity deficit}
\label{sec:cvd}
As recalled in the previous section, the coorbital vortensity deficit $\delta m$ is key to define when migration runs away. When $\delta m < M_{\rm p}$, the dynamical corotation torque is not large enough to trigger runaway migration. On the contrary, when $\delta m > M_{\rm p}$, $\mathcal{P}$ enters a runaway migration regime for which small perturbations to the position of the planet are rapidly amplified. The coorbital vortensity deficit reads \citep{Masset2003}:
\begin{equation}
\delta m = 4\pi r_{\rm p}\omega(r_{\rm p}) \times
\left[x_{\rm s}\frac{\Sigma}{\omega}\left(r_{\rm p}-x_{\rm s}\right)-
\int_{r_{\rm p}-x_{\rm s}}^{r_{\rm p}} \frac{\Sigma}{\omega}\left(r\right) dr\right],
\label{eq:deficit}
\end{equation}
with $\omega$ the gas vorticity and $x_{\rm s}$ the radial half-width of the planet's horseshoe region. The first term in the square brackets features the inverse vortensity of the gas entering the horseshoe region at orbital radius $r_{\rm p} - x_{\rm s}$, the second term the inverse vortensity of the gas trapped in libration inside the horseshoe region. Eq.~(\ref{eq:deficit}) is valid while the planet's migration timescale over a radial distance $x_{\rm s}$ remains much longer than the half-libration timescale of fluid elements at the horseshoe separatrices, located at $r_{\rm p} \pm x_{\rm s}$. In other words, $\delta m$ can be expressed by Eq.~(\ref{eq:deficit}) so long as the migration speed $\dot{a}$ is slow compared to the critical speed $\dot{a}_{\rm c}$:
\begin{equation}
|\dot{a}| \ll \dot{a}_{\rm c} = \frac{3x_{\rm s}^2 \Omega_{\rm p}}{4\pi r_{\rm p}},
\label{eq:slowspeed}
\end{equation}
with $\Omega_{\rm p}$ the orbital frequency of $\mathcal{P}$. When this is the case, the horseshoe region can be approximated as a rectangle in polar cylindrical coordinates, with a radial width of $2x_{\rm s}$ and an azimuthal width of nearly $2\pi$. We will also assume that the gas vortensity trapped in libration with $\mathcal{P}$ is uniform \citep[see, e.g.,][and Fig.~\ref{fig:invortensity} below]{Paardekooper2014}. Further denoting by $\mathcal{I_V}$ the inverse vortensity $\Sigma/\omega$, by $\mathcal{I_V}_{e}$ the inverse vortensity of the gas entering the horseshoe region ($\mathcal{I_V}_{e} = \mathcal{I_V}(r_{\rm p}-x_{\rm s})$), and by $\mathcal{I_V}_{\rm lib}$ that of the gas trapped in libration, Eq.~(\ref{eq:deficit}) can thus be recast as
\begin{equation}
\begin{aligned}
\delta m \approx 4\pi r_{\rm p}\omega(r_{\rm p})x_{\rm s} \times
\left[\mathcal{I_V}_{e}-
\mathcal{I_V}_{\rm lib}\right],
\label{eq:deficit_approx}
\end{aligned}
\end{equation}
which agrees with the simplified version of $\delta m$ obtained in \citet{Lin2010} (see their Section~6.1.1). When Eq.~(\ref{eq:slowspeed}) is no longer satisfied, both the material trapped in libration with $\mathcal{P}$ and the orbit-crossing flow have an azimuthal extent $\Delta \theta<2\pi$. Assuming that the expression of $\delta m$ in Eq.~(\ref{eq:deficit}) remains valid even when $|\dot{a}| \sim \dot{a}_{\rm c}$, we show in Appendix~\ref{sec:appendix_b} that the approximated expression for $\delta m$ in Eq.~(\ref{eq:deficit_approx}) gets multiplied by $\Delta \theta / 2\pi$ (see Eq.~\ref{eq:appendix_deficit}). A schematic illustration of $\mathcal{I_V}_{e}$ and $\mathcal{I_V}_{\rm lib}$ can be found in the top-middle panel of Fig.~\ref{fig:invortensity}.

\begin{figure}
\centering
\includegraphics[width=1.07\hsize]{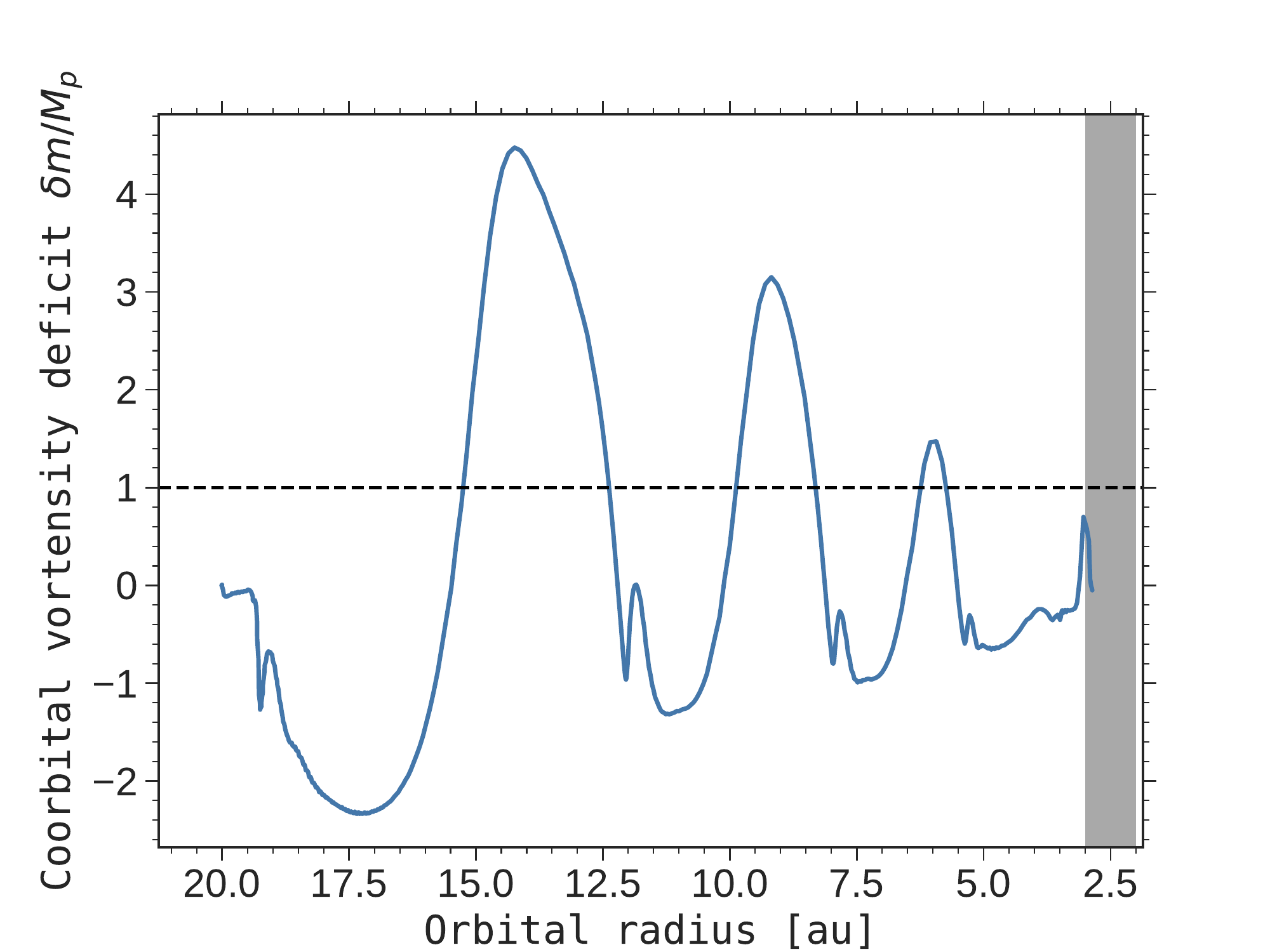}
\caption{Numerical estimation of the coorbital vortensity deficit ($\delta m$) during the inward migration of $\mathcal{P}$ in the Uther simulation. $\delta m$ is normalized by the planet's mass $M_{\rm p}$. The horizontal dashed line separates the regimes of runaway migration ($\delta m / M_{\rm p} \geq 1$) and non-runaway migration ($\delta m / M_{\rm p} \leq 1$). The gray band corresponds to the inner wave-killing zone.}
\label{fig:deficit}
\end{figure}

We display in Fig.~\ref{fig:deficit} the coorbital vortensity deficit normalized by the planet's mass $M_{\rm p}$. For simplicity, $\delta m$ is calculated via Eq.~(\ref{eq:deficit_approx}) instead of Eq.~(\ref{eq:appendix_deficit}), as an automatic determination of $\Delta \theta$ from the simulation outputs would be quite cumbersome. For $\omega(r_{\rm p})$, we have checked that it can be approximated as the unperturbed (initial) value of the gas vorticity at $r=r_{\rm p}$. For $x_{\rm s}$, we use $x_{\rm s} \approx 2.45 r_{\rm p} \times (q_{\rm p}/3)^{1/3}$, which applies to planets with $q_{\rm p}/h^3(r_{\rm p}) \gtrsim 2$ on fixed circular orbits \citep{Masset2006, Jimenez2017}. We have checked by a streamline analysis that this expression still works correctly in our simulations. A systematic determination of $\mathcal{I_V}_{e}$ and $\mathcal{I_V}_{\rm lib}$ from the simulation outputs would also be a little tricky. For $\mathcal{I_V}_{e}$, we approximate it as the inverse vortensity radially averaged over horseshoe U-turns just behind the planet in azimuth (since the planet migrates inward), using the fact that the inverse vortensity over a radial U-turn remains essentially unaltered for the range of viscosities in our disk models. This method works generally well, except at the initiation and termination of the runaway stages, where a streamline analysis would be required to estimate $\mathcal{I_V}_{e}$ more accurately. For the quantity $\mathcal{I_V}_{\rm lib}$, it is estimated as the average inverse vortensity of gas in front of the planet in azimuth around $r=r_{\rm p}$, using the fact that, as already stated below, the vortensity of the gas trapped in libration with the planet is roughly uniform. This method is less robust at the termination of runaway stages because of vortensity mixing within the horseshoe region.

\begin{figure*}
\centering
\includegraphics[width=1\hsize]{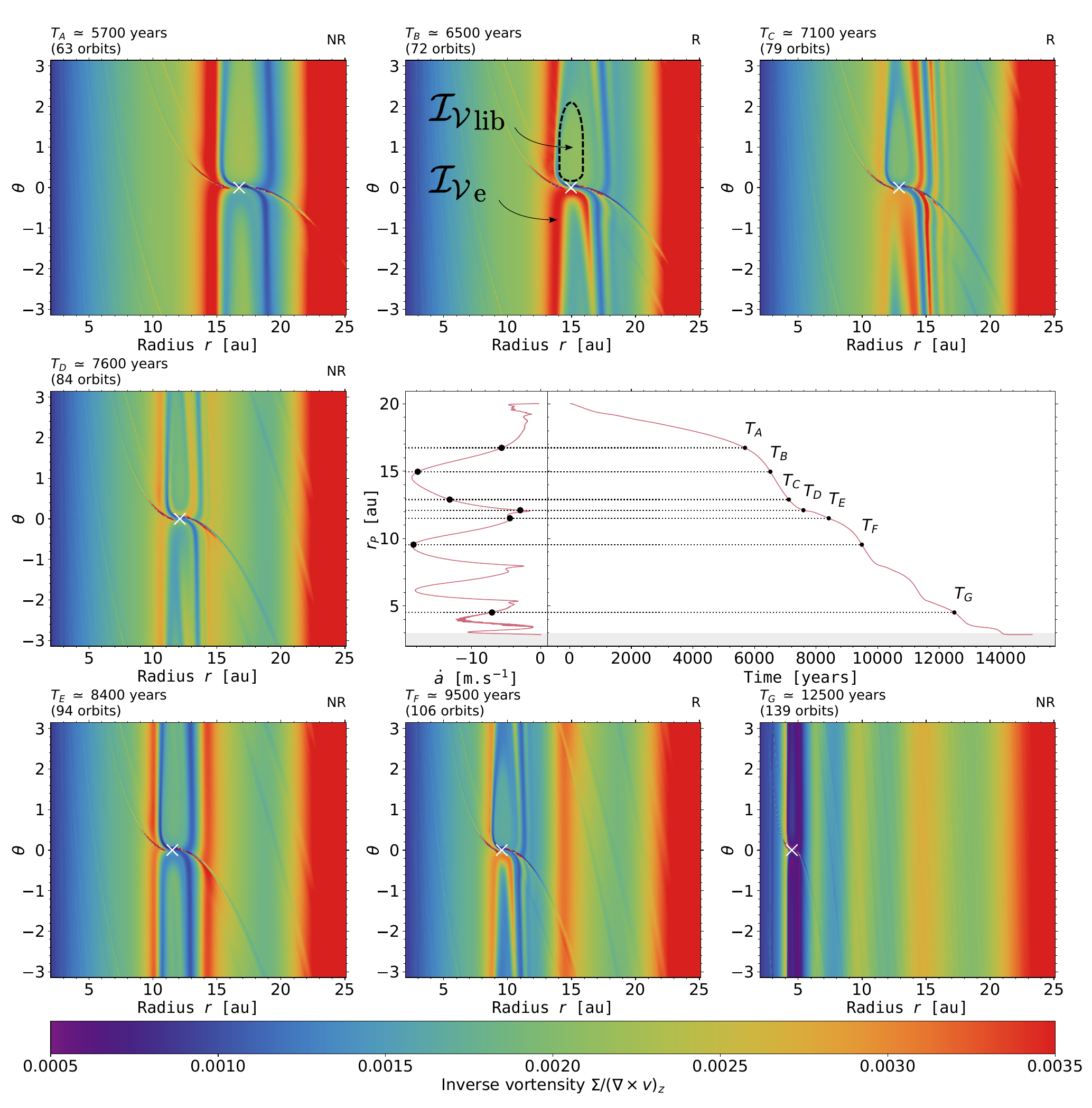}
\caption{Results of the Uther simulation: time evolution of the planet's orbital radius and migration rate (middle-right panel; the gray band corresponds to the inner wave-killing zone), and screenshots of the inverse vortensity ($\mathcal{I_V}$) at seven times denoted by $T_{\rm A}$ to $T_{\rm G}$ in the panel showing the planet's orbital evolution. In each screenshot, $\mathcal{I_V}$ is expressed in code units and shown in polar cylindrical coordinates between the inner boundary at 2 au and 25 au, and the number of orbits indicated in the upper-left corner corresponds to the number of orbital periods at the planet's initial location (20 au). Note that one orbital period at this location corresponds to $\approx$90 years. In the upper-right corner of each $\mathcal{I_V}$ panel, we specify with 'R' or 'NR' whether the planet is undergoing runaway or non-runaway migration. The planet's position is marked by a white cross.}
\label{fig:invortensity}
\end{figure*}

Fig.~\ref{fig:deficit} clearly shows that $\delta m$ oscillates around $M_{\rm p}$, which indicates that $\mathcal{P}$ alternates between stages of runaway and non-runaway migration, and which explains the steps obtained in the migration pattern (Fig.~\ref{fig:general}). Note that when the planet runs away, the material that remains in corotation (or in libration) with $\mathcal{P}$ shrinks to a trapezoidal region in the radius-azimuth plane, either ahead or behind the planet in azimuth depending on the direction of migration \citep{Masset2002,Peplinski2008b,Peplinski2008c}. The trapezoidal region, which in our case is ahead of the planet as the latter migrates inwards, has an azimuthal width $\Delta\theta < 2\pi$ when migration runs away, as will be illustrated in Fig.~\ref{fig:invortensity} (see the libration region labeled with $\mathcal{I_V}_{\rm lib}$ in the top middle panel). Therefore, $\delta m$ is actually overestimated in Fig.~\ref{fig:deficit} when $\delta m \gtrsim M_{\rm p}$ (see Eq.~\ref{eq:appendix_deficit}). But in this case, the hypothesis of slow migration (see Eq.~\ref{eq:slowspeed}) required to express $\delta m$ via Eqs.~(\ref{eq:deficit}),~(\ref{eq:deficit_approx}) or~(\ref{eq:appendix_deficit}) may not be verified in the simulation.

To understand what stops and reinstates runaway migration intermittently, we display in Fig.~\ref{fig:invortensity} seven screenshots of the gas inverse vortensity ($\mathcal{I_V}$) taken at times $T_{\rm A}$ to $T_{\rm G}$ indicated in the middle-right panel of the same figure, which shows the temporal evolution of the planet's orbital radius ($r_{\rm p}$) and migration speed ($\dot{a}$). $\mathcal{I_V}$ is calculated in the inertial frame. We now describe each panel of $\mathcal{I_V}$ separately:
\begin{itemize}
    \item $T_{\rm A}$ (top-left panel): $\mathcal{P}$ being quite massive ($q_{\rm p} / h^3(r_{\rm p,0}) \sim 1.5$), its wakes turn into shocks within their excitation region \citep{Goodman2001} and create $\mathcal{I_V}$ maxima on both sides of the planet orbit \citep{Lin2010}. (Note that the background profile of $\mathcal{I_V}$ increases as $r^{1/2}$, which reflects its initial profile. Note also the $\mathcal{I_V}$ perturbations in the planet's wakes which arise because $q_{\rm p} / h^3(r_{\rm p,0}) > 1$.) We observe an asymmetry in the radial location of the $\mathcal{I_V}$ maxima with respect to $\mathcal{P}$, which comes about because of the inward migration of $\mathcal{P}$. Around the gap's inner edge, the synodic period of the gas relative to the planet remains shorter than the migration timescale of $\mathcal{P}$ across its gap region. The inner wake of $\mathcal{P}$ can therefore shock the gas near the gap's inner edge multiple times, and sustain an $\mathcal{I_V}$ maximum there (it is like a cumulative effect). Because of the inward migration, the outer wake of $\mathcal{P}$ cannot shock the same fluid elements repetitively near the gap's outer edge, which explains the lack of an $\mathcal{I_V}$ maximum at this location. The $\mathcal{I_V}$ maximum that is visible at around 24 au was triggered by the outer wake of $\mathcal{P}$ when its migration rate was smaller near the beginning of the simulation.
    \item $T_{\rm B}$ (top-middle panel): as migration proceeds, the material that crosses the planet's orbit has increasing inverse vortensity, which therefore increases the dynamical corotation torque and the migration rate \citep[][and see Eq.~\ref{eq:deficit_approx}]{Masset2003}. As the migration rate increases, $\mathcal{P}$ enters the runaway migration regime ($\delta m \gtrsim M_{\rm p}$). At some point, migration is fast enough that the synodic period of the gas near the gap's inner edge becomes longer than the migration timescale of $\mathcal{P}$ across its gap. The cumulative shocks of the inner wake brought about above no longer operate, or, said differently, the inner wake of $\mathcal{P}$ can no longer sustain an $\mathcal{I_V}$ maximum inside the orbit. When the entering inverse vortensity takes its maximum value, $\mathcal{P}$ reaches its maximum migration rate. This is the case shown in this panel (see the "red" stream of high $\mathcal{I_V}$ executing an outward U-turn relative to $\mathcal{P}$ at $\theta \lesssim 0$). Meanwhile, the azimuthal extent $\Delta\theta$ of the libration region decreases, which saturates the runaway process by limiting the increase in $\delta m$ (see the $\Delta\theta$ term in Eq.~\ref{eq:appendix_deficit}).
    \item $T_{\rm C}$ (top-right panel): after the gas with maximum $\mathcal{I_V}$ has crossed the orbit, $\mathcal{P}$ keeps on migrating and now forces material with unperturbed (and thus smaller) $\mathcal{I_V}$ to execute outward U-turns. The feedback on migration due to the coorbital vortensity deficit now acts in the opposite way: the inverse vortensity of the material entering the horseshoe region ($\mathcal{I_V}_{e}$) decreases, $\delta m$ decreases, and so does the migration rate.
    \item $T_{\rm D}$ (middle-left panel): migration is now slow enough that part of the high $\mathcal{I_V}$ material that previously entered the horseshoe region becomes trapped in libration and performs multiple horseshoe U-turns. This panel actually highlights a secondary inward U-turn of high $\mathcal{I_V}$ material in front of the planet in azimuth. Because the gas executing inward U-turns has a higher $\mathcal{I_V}$ than the gas doing outward U-turns, a positive corotation torque applies to $\mathcal{P}$, which slows down even more its migration, until the vortensity in the horseshoe region gets progressively mixed. These secondary U-turns will be examined in more detail in Section~\ref{sec:impact_vort}.
    \item $T_{\rm E}$ (bottom-left panel): $\mathcal{P}$ is now in a non-runaway migration regime and its wakes can trigger new $\mathcal{I_V}$ maxima on both sides of the planet's orbit. Inward migration resumes, and a radial asymmetry in the position of the $\mathcal{I_V}$ maxima with respect to $\mathcal{P}$ can be seen again in this panel, which is similar to the one shown at time $T_{\rm A}$. A new cycle of runaway migration begins. The comparison with the panel at $T_{\rm A}$ highlights that the radial location of the $\mathcal{I_V}$ maximum outside the planet at $T_{\rm E}$ ($\sim 14$ au) nearly coincides with that of the $\mathcal{I_V}$ maximum inside the planet at $T_{\rm A}$, i.e. just before the onset of runaway migration. We will come back to this point in Section~\ref{sec:ccl}.
    \item $T_{\rm F}$ (bottom-middle panel): as previously, runaway stalls as the inner wake ultimately fails to maintain an $\mathcal{I_V}$ maximum inside the planet's orbit and the azimuthal width of the libration region shrinks. This panel is very similar to the one shown at time $T_{\rm B}$, and we notice that the migration rate takes very similar values at both times. A new maximum of $\mathcal{I_V}$ will form outside the planet's orbit when the migration rate has reached again its minimum value.
    \item $T_{\rm G}$ (bottom-right panel): at the end of the simulation, when migration stalls near the grid's inner boundary, multiple bands of high-$\mathcal{I_V}$ material are formed, including one inside the planet's orbit. The bands beyond the planet's orbit trace the transition stages between fast and slow migration. The longevity of these bands will be discussed in Sections~\ref{sec:dsd_1} and~\ref{sec:longevity}.
\end{itemize}

As already stated in Section~\ref{sec:overview}, \citet{Lin2010} and \citet{McNally2019} have obtained a similar scenario of intermittent migration. Although in these models gas vortices form at the edges of the planet gap due to the Rossby-Wave Instability, the intermittent episodes of runaway migration can also be accounted for by the time evolution of the inverse vortensity of the gas crossing the planet's orbit.

\begin{figure*}
\begin{center}
\includegraphics[width=1\hsize]{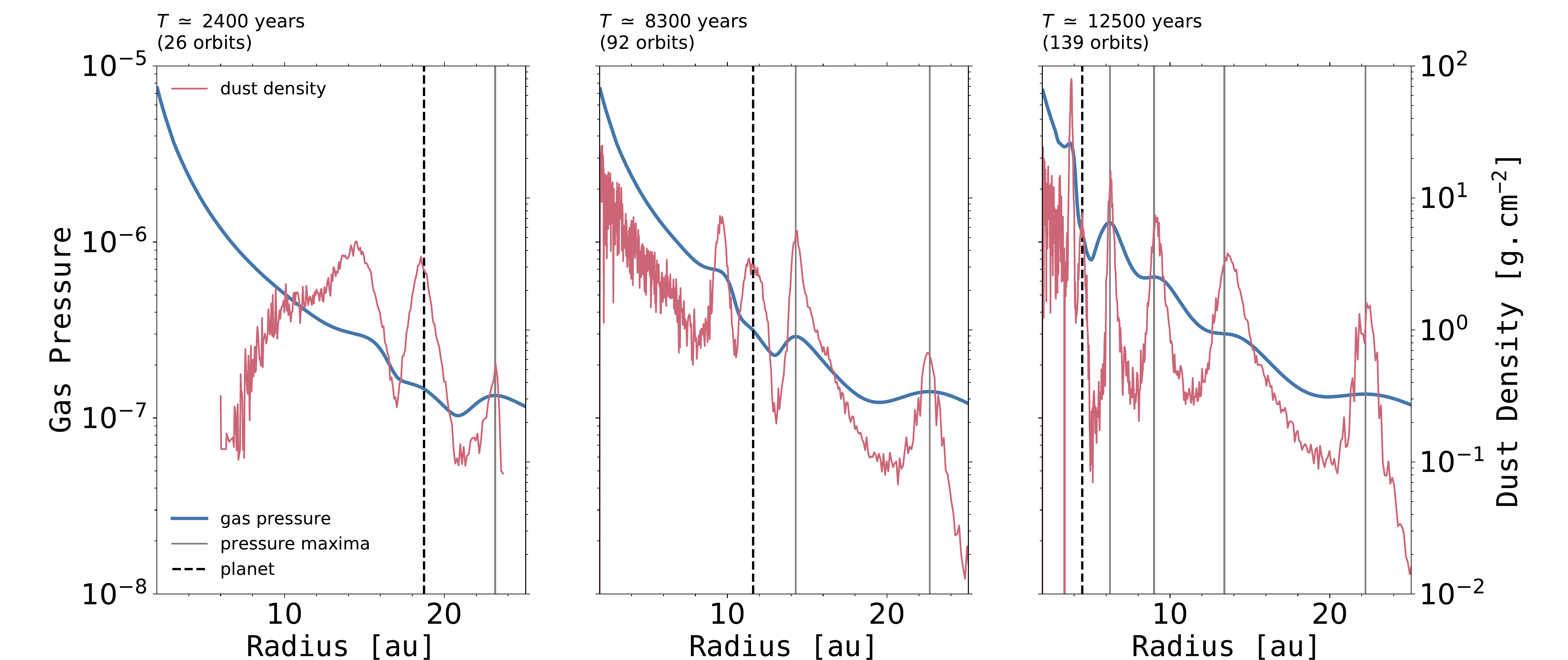}
\includegraphics[width=1\hsize]{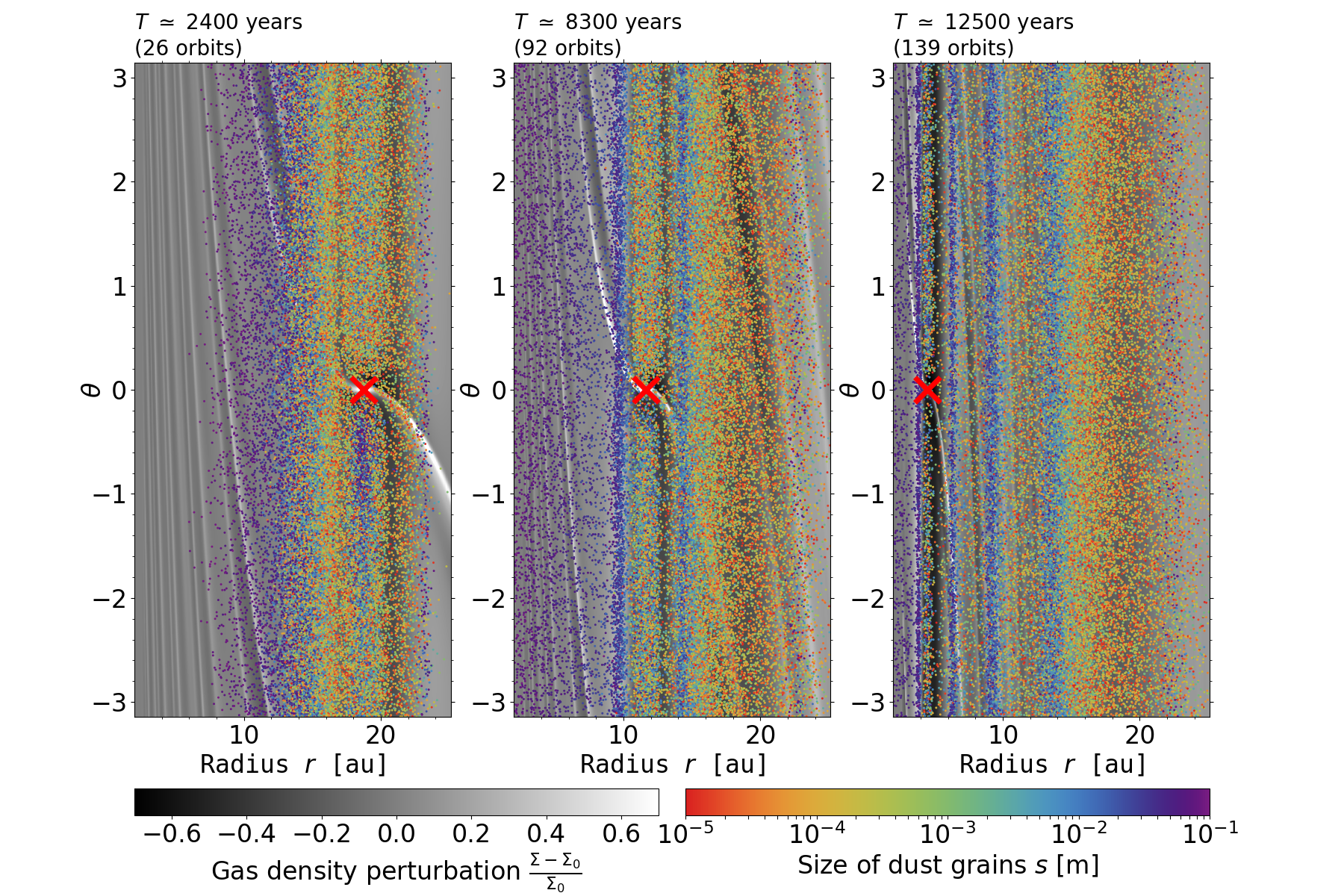}
\end{center}
\caption{Results of the Uther simulation: gas structure and dust spatial distribution in polar cylindrical coordinates in the disk region between the inner boundary at 2 au and 25 au. Results are shown at 2400 years in the left panels, at 8300 years in the middle panels, and at 12500 years in the right panels (same time as in Fig.~\ref{fig:invortensity}, bottom right). \textbf{Top:} azimuthally-averaged radial profiles of the gas pressure (\textit{blue}) and of the dust's surface density (\textit{red}). The planet's orbital radius is spotted by a vertical dashed line. Pressure maxima outside the planet position are depicted by the thick gray lines. \textbf{Bottom:} perturbed gas surface density relative to its initial profile, ($\Sigma-\Sigma_0)/\Sigma_0$, in black and white. The location of dust particles is overlaid by colored dots. The rainbow color bar shows dust size in meters, with purple dots for the largest particles ($10$ cm) and red dots for the smallest ($10~\mu$m). In each panel, the red cross at zero azimuth marks the position of the planet. }
\label{fig:denspress3}
\end{figure*}

\subsubsection{Dust spatial distribution}
\label{sec:dsd_1}
We have seen in Section~\ref{sec:cvd} that intermittent planet migration gives structure to the disk gas, with the formation of new maxima in the inverse vortensity profile on each side of the planet's orbit at the end of each stage of runaway migration. These inverse vortensity maxima are intimately related to pressure maxima triggered by the planet wakes when migration decelerates, as we illustrate below (the reader is referred to, e.g., Eq.~32 in \citealp{Casoli2009} for a formal link between maxima of inverse vortensity and of pressure). We then study how intermittent migration impacts the dust distribution. For the simulation Uther, we use $5 \times 10^4$ Lagrangian test particles with initial orbital radius $r_{\rm d} \in \left[14-21\right]$ au. This is meant to maximize the particles resolution at the dust structures that form during the inward migration of $\mathcal{P}$.

The upper panels in Fig.~\ref{fig:denspress3} show three snapshots of the azimuthally-averaged radial profiles of the gas pressure (in blue) and of the dust's surface density (in red). The dust's surface density is calculated with the dust's size distribution and total mass adopted in the radiative transfer calculations (see Section~\ref{sec:RTsetup}). We note in passing that the dust-to-gas surface density ratio does not exceed $\sim$0.1 throughout the simulation, which justifies that dust feedback on the gas can be safely discarded for the Uther disk model. The thick gray lines in Fig.~\ref{fig:denspress3} show the location of the pressure maxima beyond the planet's orbital radius. The location of the planet is marked by the dashed line.

Shortly after the beginning of the simulation (Fig.~\ref{fig:denspress3}, top left), there is only one pressure maximum at the outer edge of the planet's gap, which is built up by the outer wake of $\mathcal{P}$ before the first episode of runaway migration (see also the Appendix A of \citealp{Meru2019}). At the end of the first cycle of runaway migration, when $\mathcal{P}$ has decelerated, a new pressure maximum forms outside the planet orbit (see Fig.~\ref{fig:denspress3}, top middle, which corresponds approximately to Fig.~\ref{fig:invortensity}, bottom left, for the inverse vortensity). At the end of the third episode of runaway migration (Fig.~\ref{fig:denspress3}, top right, corresponding to Fig.~\ref{fig:invortensity}, bottom right), we obtain four pressure maxima outside the planet orbit, and one inside the orbit. (The fact that the latter is not an inflection point may be partly due to the proximity of the inner wave-killing zone). More precisely, at the position of $\mathcal{P}$ in this last snapshot ($\sim$5 au), the planet braked three times, which gave it time to create three pressure maxima in addition to the initial pressure maximum outside the original position of $\mathcal{P}$ (near 22 au). The pressure maxima that are not sustained by the planet wakes progressively turn into inflection points under the action of turbulent viscosity.

At a given location in the disk, dust and gas do not have the same orbital speed due to the pressure-gradient force acting on the gas. This velocity difference induces an aerodynamic force from the gas to the dust, which eventually leads to the radial drift of the dust. At a pressure maximum, there is no relative motion of the dust and the gas, therefore no aerodynamic force and no resulting drift. A pressure maximum thus defines an equilibrium position where dust accumulates and is trapped in rings. We thus expect an accumulation of dust at each bump in the pressure profile, and this is precisely what can be seen in the upper panels in Fig.~\ref{fig:denspress3} by comparing the location of the maxima in the dust surface density (red curves) and in the gas pressure (thick gray lines) beyond the planet.

The bottom panels of Fig.~\ref{fig:denspress3} display in black and white the perturbed gas surface density ($\Sigma - \Sigma_0$) relative to its initial profile $\Sigma_0$ at the same times as the top panels in the same figure. The red cross spots the location of $\mathcal{P}$. We depict dust super-particles as colored dots; color varies with particles size, from red ($10~\mu$m) to purple ($10$ cm). The inward drift of the largest ($s \gtrsim 1$ cm) particles is clearly visible between the bottom left and bottom middle panels of Fig.~\ref{fig:denspress3}. Note that the small particles are not present in the inner disk due to the choice of initial particles location and the fact that there has not been enough time for the small particles to drift. In the bottom-right panel of Fig.~\ref{fig:denspress3}, we see that dust particles are trapped in axisymmetric rings at the location of the pressure maxima (around 22, 14, 9, 6 and 4 au for this simulation at 12500 years), and that larger particles get more concentrated at these locations on account of their larger Stokes number (larger particles, with Stokes number $\lesssim 1$, drift more rapidly towards equilibrium locations).

\begin{figure}
\centering
\resizebox{1.05\hsize}{!}
{
\hspace{-0.7cm}
\includegraphics[width=1.3\hsize]{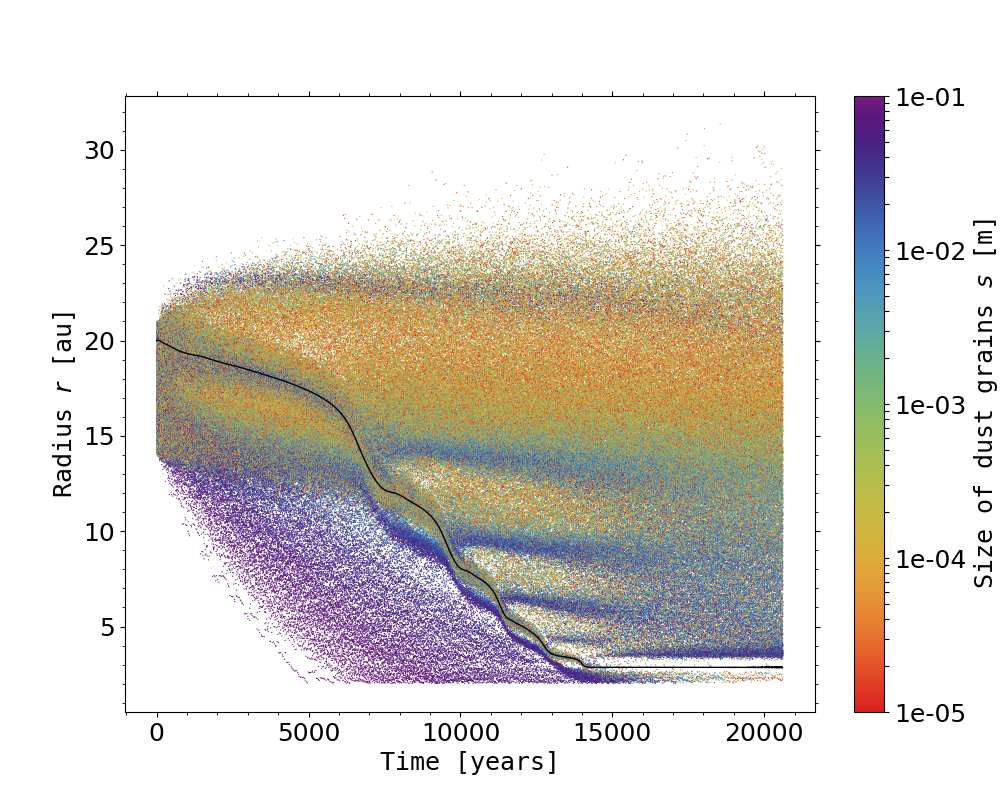}
\hspace{0.7cm}
}
\caption{Space-time diagram for the Uther simulation, illustrating the longevity of dust rings formed by multiple stages of runaway planet migration. This panel displays the orbital radius of a sample of super-particles as a function of time, with color varying with particles size. The planet's orbital radius is shown by a black curve.}
\label{fig:spacetime_uther}
\end{figure}

To examine the longevity of the dust rings, we display in Fig.~\ref{fig:spacetime_uther} the space-time diagram of the dust's radial location versus dust size for the Uther simulation. Each dot marks the orbital radius of a sample of super-particles as a function of time, with color varying with particles size (see color bar on the right-hand side). This graph clearly shows the progressive disappearance of the dust rings formed prior to each stage of inward runaway migration. Dust rings typically last between 3000 and 10000 years. It is also instructive to examine the time evolution of the radial location of the largest particles (dark-blue dots). Some of these large particles are initially located around the planet's orbital radius before the first episode of runaway migration at $\sim$6000 years, and going back to the bottom-left panel in Fig.~\ref{fig:denspress3} we see that these correspond to the particles trapped around the L5 Lagrange point behind the planet in azimuth. As the runaway proceeds, these particles escape the Lagrange point and a large fraction of them become trapped at the pressure maximum outside the planet's orbit which forms at the end of this first runaway. Similarly, close inspection at the second stage of runaway migration at $\sim$9000 years shows that the large particles that become trapped at the newly formed pressure maximum outside the planet's orbit mainly originate from the reservoir of large particles initially inside the orbit. A similar observation can be made with the subsequent stages of runaway migration. It is interesting to realize that not all the reservoir of large particles inside the orbit ends up populating the dust trap triggered outside the orbit. Part of the particles embarking onto outward U-turns perform secondary U-turns inward, the same secondary U-turns that we have illustrated on the gas inverse vortensity in the middle left panel ($T_{\rm D}$) of Fig.~\ref{fig:invortensity}. This process allows to maintain a contingent of large particles inside the planet's orbit which will progressively fill the dust traps outside the planet's orbit formed after each jolt of the planet. We finally notice that the dust traps formed outside the planet's orbit seem to have a geometrical radial spacing (there is a factor $\sim 1.5$ between the radial locations of successive dust rings, we will come back to this point in Section~\ref{sec:ccl}).

\begin{figure}
\centering
\resizebox{0.94\hsize}{!}
{
\includegraphics[width=0.99\hsize]{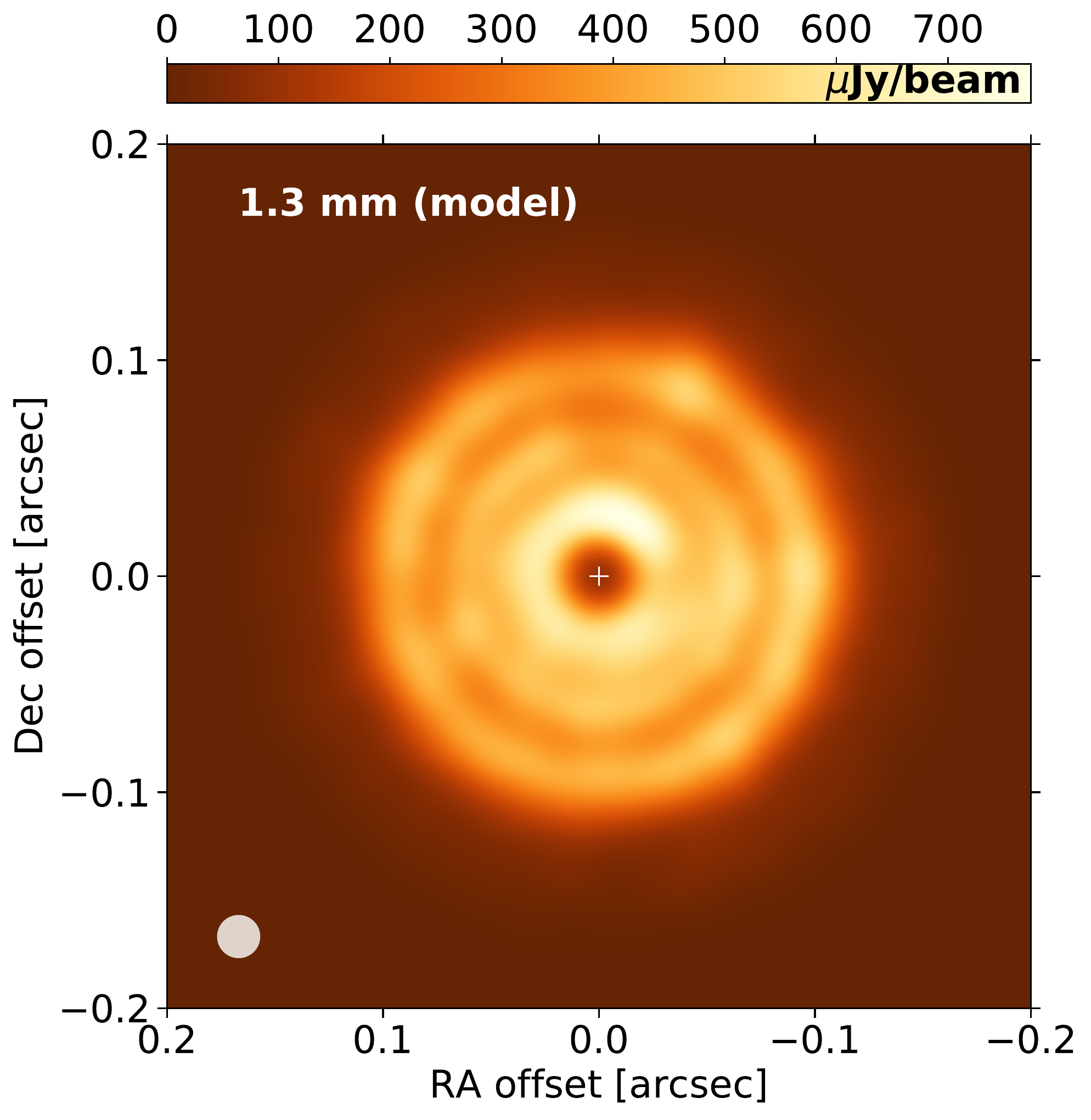}
}
\resizebox{0.94\hsize}{!}
{
\includegraphics[width=0.99\hsize]{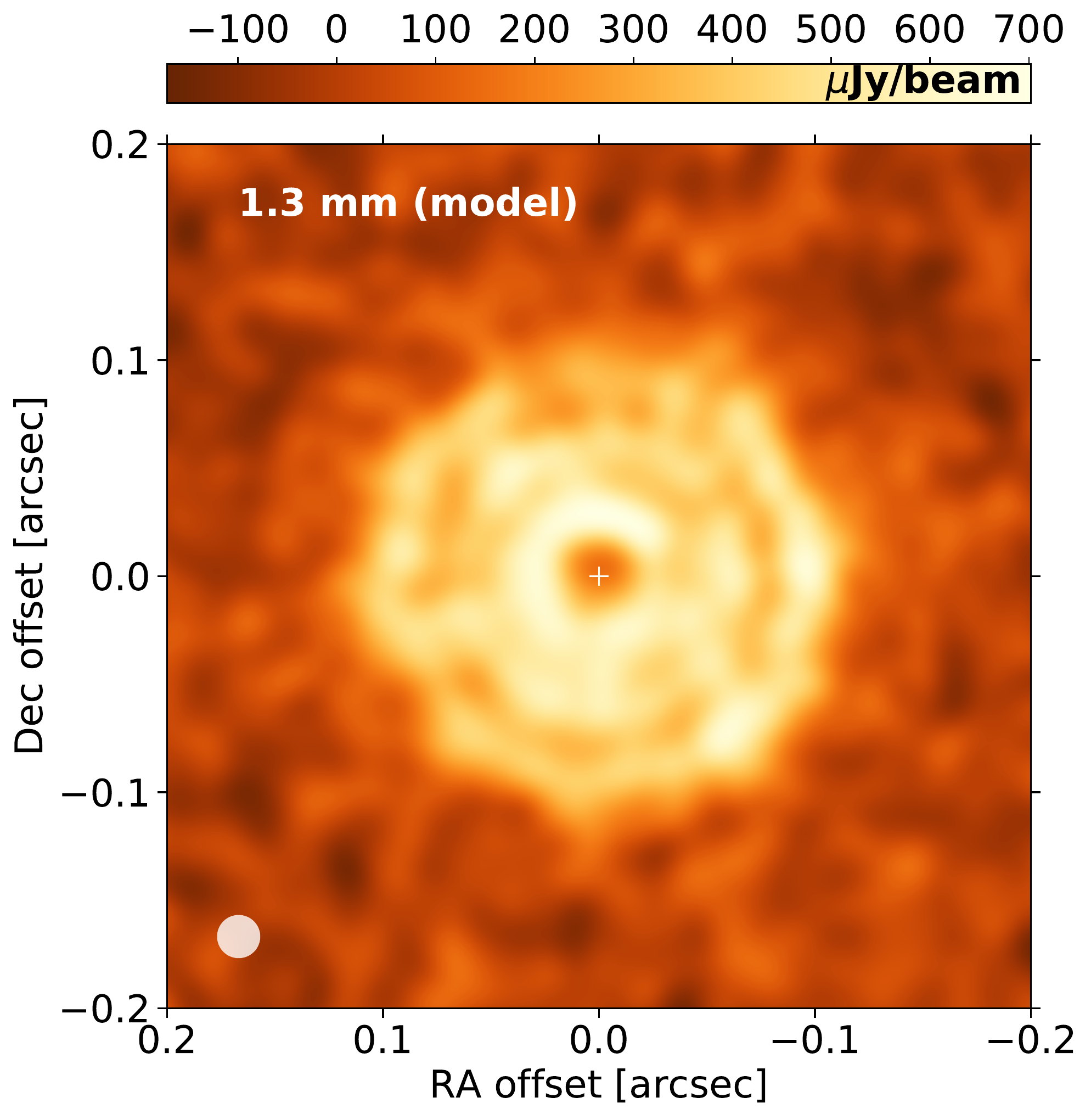}
}
\resizebox{0.92\hsize}{!}
{
\hspace{0.2cm}
\includegraphics[width=0.99\hsize]{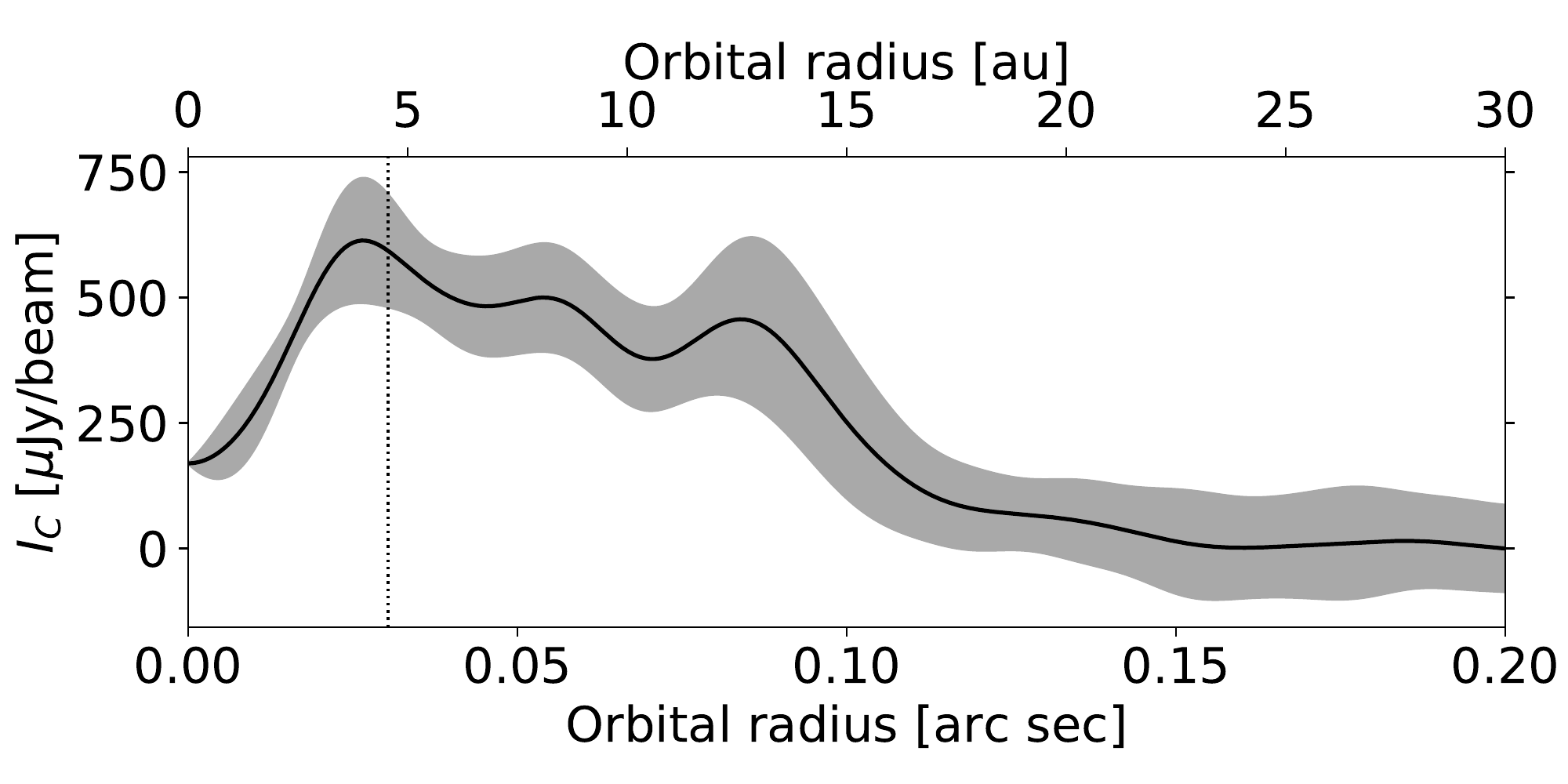}
}
\caption{Predicted continuum emission at 1.3 mm for the Uther simulation at 12500 years (or 139 planet orbits). \textbf{Top:} Synthetic flux map obtained after convolution with a $0\farcs02$ circular beam. The beam is shown by the circle in the bottom-left corner. The $x$- and $y$-axes indicate the offset from the stellar position in the RA and DEC in arcseconds, i.e., north is up and west is to the right. \textbf{Middle:} Same as upper panel, but with white noise of 50 $\mu$Jy/beam standard deviation added to the raw flux maps (prior to beam convolution). \textbf{Bottom:} Azimuthally averaged convolved intensity as a function of distance to the star. The gray shaded area around the profile corresponds to $\pm 2 \sigma$ dispersion around the mean. The vertical dotted line shows the location of the planet.}
\label{fig:uter_rt}
\end{figure}

\subsubsection{Continuum emission at $\lambda=1.3$ mm}
\label{sec:dte_1}
The dust spatial distribution obtained in the Uther simulation was post-processed with the 3D radiative transfer code RADMC3D to produce synthetic flux maps of continuum emission at $\lambda = 1.3$ mm according to the procedure detailed in Section~\ref{sec:RTsetup}. The main parameters of the dust radiative transfer calculations are summed up in Table~\ref{table:rt_parameters}.

\begin{table}
\centering
\caption{\label{table:rt_parameters} Parameters of the dust radiative transfer calculations}
\begin{tabular}{lr}
\hline
\hline
Parameter                           		& Value           \\
\hline
Wavelength                           		& 1.3 mm      \\
Dust size range           		& $s \in [10^{-5} - 10^{-1}]$ m      \\
Dust size distribution           	& $n(s) \propto s^{-3.5}$   \\
Dust-to-gas mass ratio           	& $10^{-2}$  \\
Dust composition           		& 60\% silicates, 40\% water ices\\
Scattering                        		& discarded   \\
Disk distance                        	& 150 pc       \\
Stellar radius                        		& 2 R$_{\odot}$       \\
Star effective temperature           & 7000 K       \\
\hline
\end{tabular}
\end{table}

The synthetic flux map of continuum emission based on the results of the Uther simulation at 12500 years is displayed in the upper panel in Fig.~\ref{fig:uter_rt}. Three bright rings of emission are clearly visible in the flux map. The outermost visible ring, which is located at $\sim0\farcs09$ ($\sim14$ au), corresponds to the maximum in the dust's surface density located near 14 au in the top-right panel in Fig.~\ref{fig:denspress3}. The dust density maximum located near 22 au does not produce a visible bright ring in the flux map. The second visible ring in the upper panel of Fig.~\ref{fig:uter_rt} mostly corresponds to the dust density maximum located near 9 au in the top-right panel in Fig.~\ref{fig:denspress3}. Finally, the innermost bright ring corresponds to the beam dilution of the three dust density maxima which are located at the planet's orbit radius and on both sides of the planet's gap. Furthermore, we notice azimuthal variations of the flux along the bright rings, which are of the order of 50 $\mu$Jy/beam. We have checked that, at the location of the bright rings, the dust's thermal emission is optically thick (the absorption optical depth varies between about 2 and 4). These azimuthal variations in the intensity thus convey azimuthal variations in the dust temperature obtained by the thermal Monte-Carlo calculation.

The middle panel in Fig.~\ref{fig:uter_rt} shows the same flux map obtained by adding white noise to the raw flux map prior to beam convolution (see Section~\ref{sec:RTsetup}). It is clear that the three bright rings are still discernible despite the realistic level of noise that we consider. These rings are also clearly visible in the lower panel of Fig.~\ref{fig:uter_rt}, where we display the azimuthally-averaged radial profile of the convolved intensity.

\subsection{Pendragon simulation ($h_0 = 0.06$, $\alpha = 10^{-4}$)}
\label{sec:case2}

\begin{figure*}
\begin{center}
\includegraphics[width=0.99\hsize]{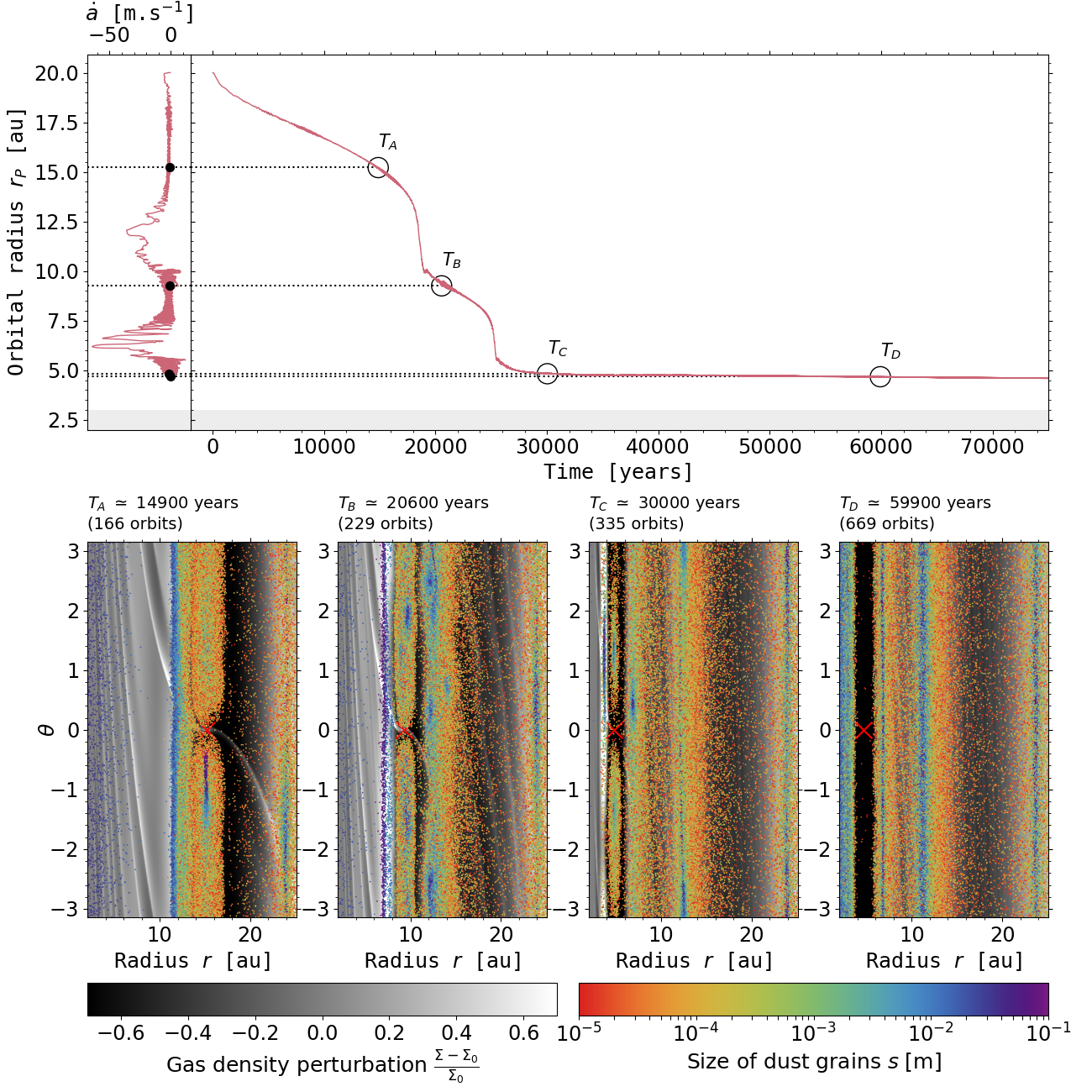}
\end{center}
\caption{Results of the Pendragon simulation : planet's orbital evolution, gas structure and dust spatial distribution in the disk region between the inner boundary at 2 au and 25 au. \textbf{Top right:} evolution of the planet's orbital radius as a function of time. The four circles indicate four different times in the simulation : $T_{\rm A} = 14900$, $T_{\rm B} = 20600$, $T_{\rm C} = 30000$ and $T_{\rm D} = 59900$ years. \textbf{Top left:} evolution of the planet's migration speed as a function of its position in the disk. The gray band corresponds to the inner wave-killing zone. \textbf{Bottom:} same as lower panel of Fig.~\ref{fig:denspress3}, but for the four aforementioned times in the Pendragon simulation.}
\label{fig:pendragon_case}
\end{figure*}

We now present the results of the Pendragon simulation, which differs from the Uther simulation by a slightly increased disk's aspect ratio ($h_0 = 0.06$ instead of 0.05) and a reduced alpha turbulent viscosity ($\alpha = 10^{-4}$ instead of $10^{-3}$). Having a smaller $\alpha$ implies that the structures that we obtain in the dust spatial distribution are sharper and less subject to turbulent diffusion, which leads to structures that last longer in the disk. However, because a low turbulence favors the Rossby wave instability (RWI), the formation of vortices that can interact with $\mathcal{P}$ makes the interpretation of the planet's orbital evolution a priori harder. These vortices can appear for example when a stream of high $\mathcal{I_V}$ material mixes with low $\mathcal{I_V}$ non-perturbed regions, which is typical of the fast migration stages presented in Section~\ref{sec:cvd}. Nevertheless, the important quantities presented in that same section, mainly the coorbital vortensity deficit and the inverse vortensity of the orbit-crossing material, allow a good understanding of what happens in the low turbulent case that we present in this section. For the dust, we use this time $10^5$ Lagrangian test particles initially distributed in a larger band $r_{\rm d} \in \left[14-23\right]$ au, in order to have more particles trapped at the outermost pressure maximum formed by the planet shortly after the beginning of the simulation. Table~\ref{table:case2_parameters} sums up the main parameters of this simulation.

\begin{table}
\centering
\caption{\label{table:case2_parameters}Parameters of simulation Pendragon (Section~\ref{sec:case2})}
\begin{tabular}{lr}
\hline
\hline
Parameter                           & Value           \\
\hline
Planet to primary mass ratio $q_{\rm p}$                  & $4.3\times10^{-4}$         \\
Disk's aspect ratio $h_0$ at 10 au                      & 0.06       \\
Alpha turbulent viscosity         & $10^{-4}$   \\
Gas surface density $\Sigma_0$ at 10 au [code units]      & $10^{-3}$   \\
\hline
Number of dust super-particles  & 100000\\
Dust's initial location  & $\in [14-23]$ au\\
\hline
\end{tabular}
\end{table}

\subsubsection{Dust spatial distribution}
\label{sec:dsd_2}

The upper panel in Fig.~\ref{fig:pendragon_case} shows the time evolution of the planet's orbital radius (right-hand side) and its migration speed (left-hand side). The migration pattern is similar to that in the Uther simulation, with a succession of fast (runaway) and slow (non-runaway) migration episodes. Compared to the Uther simulation, the steps occur at a later time because of a slower initial migration due to a deeper gap. Also, the steps are less numerous and more pronounced, foretelling there should be less bright rings in the continuum emission.

In the lower part of Fig.~\ref{fig:pendragon_case} is shown the dust spatial distribution (colored dots) superimposed on the perturbed gas surface density (black and white contours) at the four times marked by black circles in the upper panel: $T_{\rm A} = 14900$ years, $T_{\rm B} = 20600$ years, $T_{\rm C} = 30000$ years and $T_{\rm D} = 59900$ years. Initially, $\mathcal{P}$ migrates slowly. Due to this slow migration, the dust structures in the disk follow the planet's orbital motion. We can highlight two important behaviors: first, $\mathcal{P}$ forms a dust ring ($\mathcal{R}_p$) centered on its orbital radius with an accumulation of particles around the Lagrange point L5 behind the planet in azimuth. The ring $\mathcal{R}_p$ is surrounded by two rings: an internal ring $\mathcal{R}_i$ and an external ring $\mathcal{R}_{e,1}$. Secondly, we observe the rapid inward drift of the largest grains. While $\mathcal{P}$ migrates inward gently, it gets closer to $\mathcal{R}_i$ as it pushes particles inwards. Particles that were initially trapped at $\mathcal{R}_{e,1}$ drift slowly because the location of the corresponding pressure maximum does not change much. This progressively leads to a radial asymmetry of $\mathcal{R}_i$ and $\mathcal{R}_{e,1}$ compared to $\mathcal{R}_p$ \citep{Perez2019,Meru2019}. For instance, at time $T_{\rm A}$, $\mathcal{R}_i$, $\mathcal{R}_p$ and $\mathcal{R}_{e,1}$ are located at around 12 au, 15 au and 24 au, respectively. As we have seen in Section~\ref{sec:cvd}, this asymmetry enables material with higher $\mathcal{I_V}$ to enter the coorbital region, making $\mathcal{P}$ move faster inward. When the fast migration regime is triggered, $\mathcal{P}$ overtakes the internal ring $\mathcal{R}_i$ at around 10 au, and part of the dust that was trapped at $\mathcal{R}_i$ and at the Lagrange points perform a horseshoe U-turn. These particles then follow a complex dynamics due to the presence of small vortices arising because of the Rossby-Wave Instability, and form a new outer dust ring $\mathcal{R}_{e,2}$ at 12 au (see second lower panel from the left at time $T_{\rm B}$). The same behavior happens again during each jolt in the planet migration. During this evolution, the external rings $\mathcal{R}_{e}$ that are left behind $\mathcal{P}$ do not drift significantly. After the N$^{\rm th}$ jump, N+1 outer dust rings have been created. Since two stages of runaway migration occur in the simulation, three dust rings form outside the planet's location (here at around 7, 12 and 24 au).

Each jolt is associated with an increase in the migration speed (Fig.~\ref{fig:pendragon_case}, top left): between $T_{\rm A}$ and $T_{\rm B}$, and between $T_{\rm B}$ and $T_{\rm C}$. The rather large amplitude oscillations in the planet's migration speed are most probably due to the formation of several vortices that carry away sporadically high $\mathcal{I_V}$ material from the inner to the outer edges of the planet's gap ($T_{\rm B}$, see \citealp{Lin2010}). These vortices interact with the planet before merging and forming a single large-scale vortex ($T_{\rm C}$) in which dust particles larger than about a millimeter are trapped. The vortex gradually disappears and becomes an axisymmetric dust ring. At the time $T_{\rm D}$ shown in the lower-right panel of Fig.~\ref{fig:pendragon_case}, which is a long time after the planet has stalled its migration upon reaching the inner wave-killing zone, the three dust rings are still present. The external ring created last ($\mathcal{R}_{e,3}$ at 7 au) is located at the outer edge of the final planet's position $\mathcal{R}_p \sim 5$ au.

\begin{figure}
\centering
\resizebox{1.05\hsize}{!}
{
\hspace{-0.7cm}
\includegraphics[width=1.3\hsize]{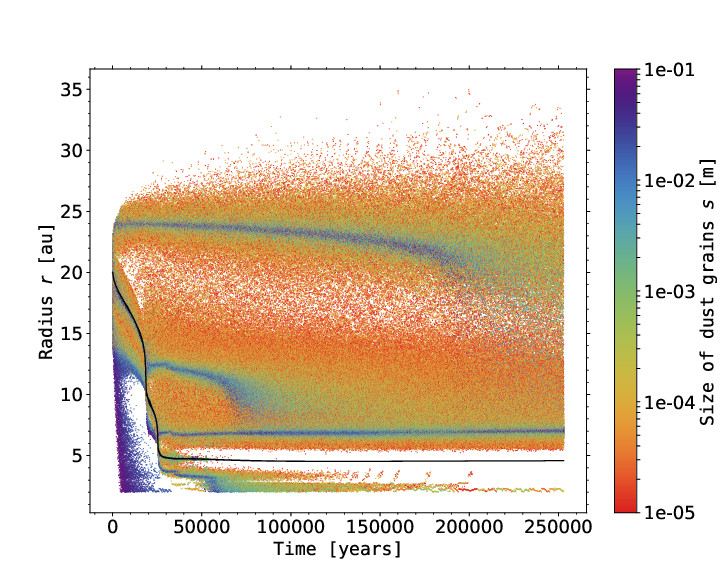}
\hspace{0.7cm}
}
\caption{Same as Fig.~\ref{fig:spacetime_uther} for the Pendragon simulation.}
\label{fig:spacetime_pendragon}
\end{figure}

We display in Fig.~\ref{fig:spacetime_pendragon} the space-time diagram for the Pendragon simulation, which shows that the dust rings have a much longer lifetime than in the Uther simulation, since some dust rings are maintained over more than $10^5$ years. We notice that the dust ring at $\sim 12$ au gets disrupted at $\sim 60000$ years onward and the largest dust particles that were initially trapped there start drifting inward. They ultimately reach the dust trap at $\sim 7$ au. Similarly, the dust ring initially at $\sim 24$ au disappears after $\sim 200000$ years. As will be further emphasized in Section~\ref{sec:longevity}, the longevity of the dust rings is intimately related to the sharpness of the pressure profile around the maxima, which turbulent viscosity progressively smooths out. It therefore depends on how strong the shock of the outer planet's wake is initially during each stage of non-runaway migration. The shorter this stage, the weaker the shock will be, and the earlier the dust ring formed beyond the planet's orbit will disappear. We finally point out that, for the size distribution and total mass of dust in the radiative transfer calculation, the dust-to-gas surface density ratio can reach up to 0.5-0.6 throughout the Pendragon simulation, and discarding dust feedback remains a safe approximation for this disk model.

\begin{figure}
\centering
\resizebox{0.94\hsize}{!}
{
\includegraphics[width=0.99\hsize]{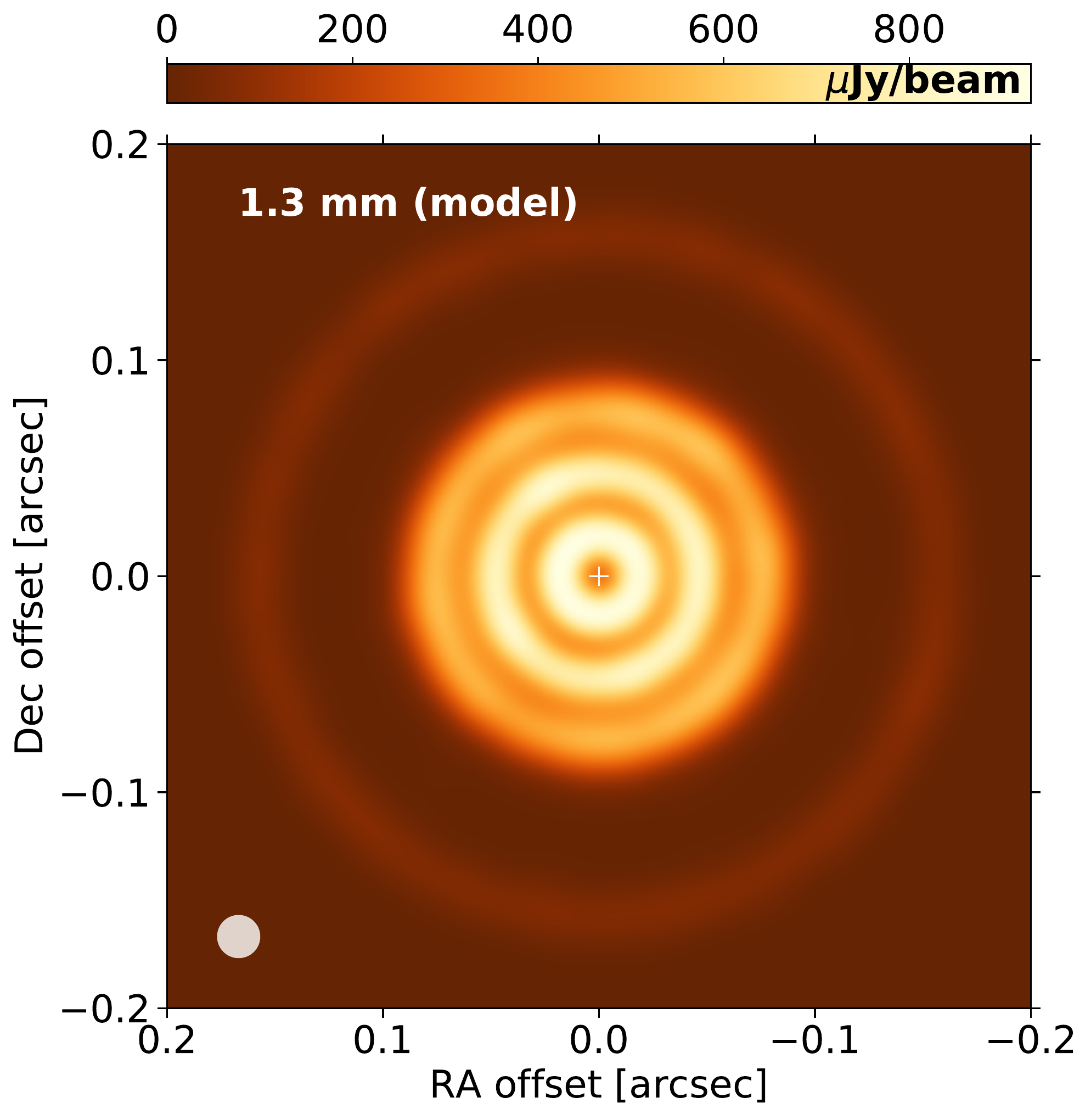}
}
\resizebox{0.94\hsize}{!}
{
\includegraphics[width=0.99\hsize]{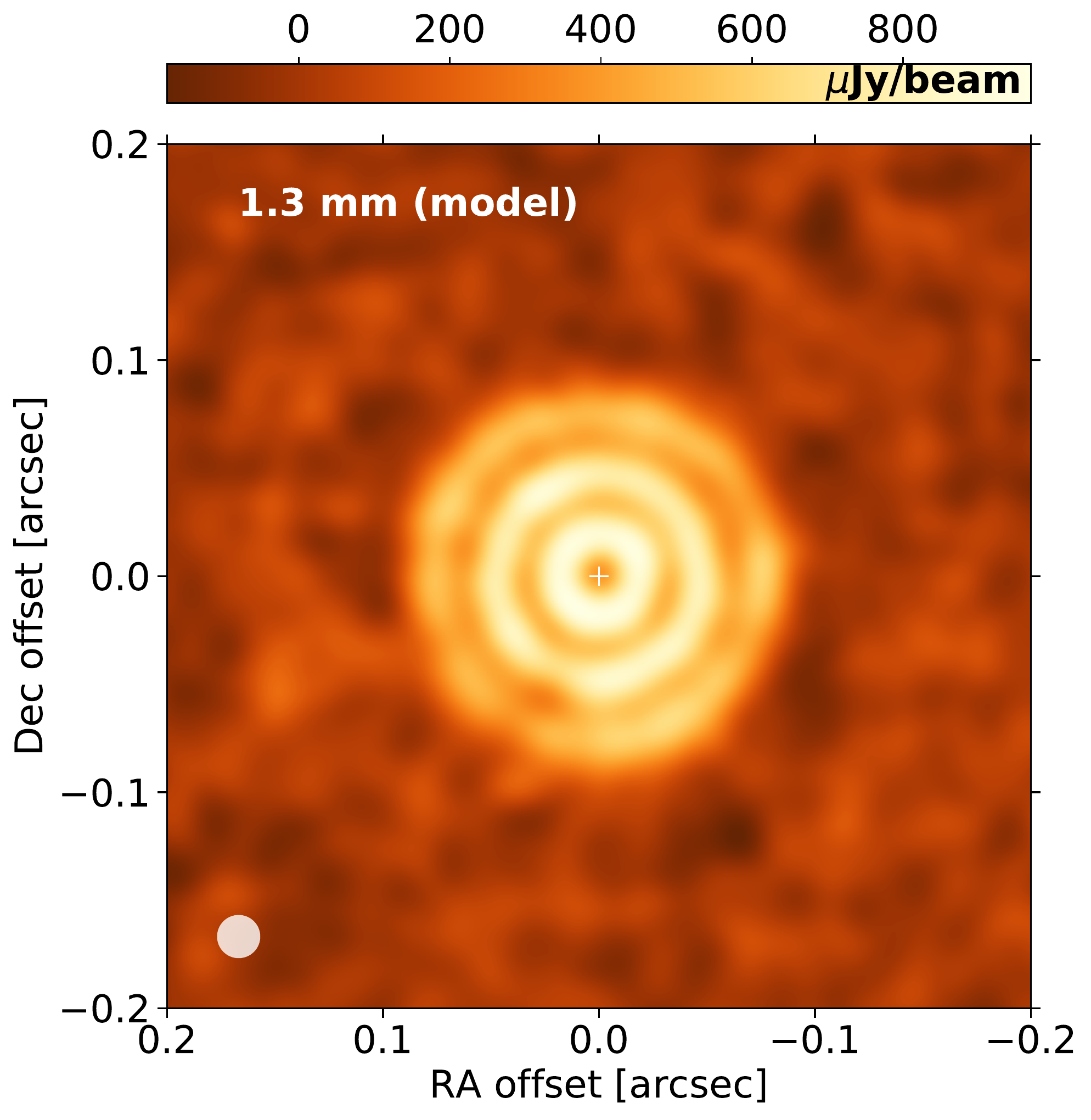}
}
\resizebox{0.92\hsize}{!}
{
\hspace{0.2cm}
\includegraphics[width=0.99\hsize]{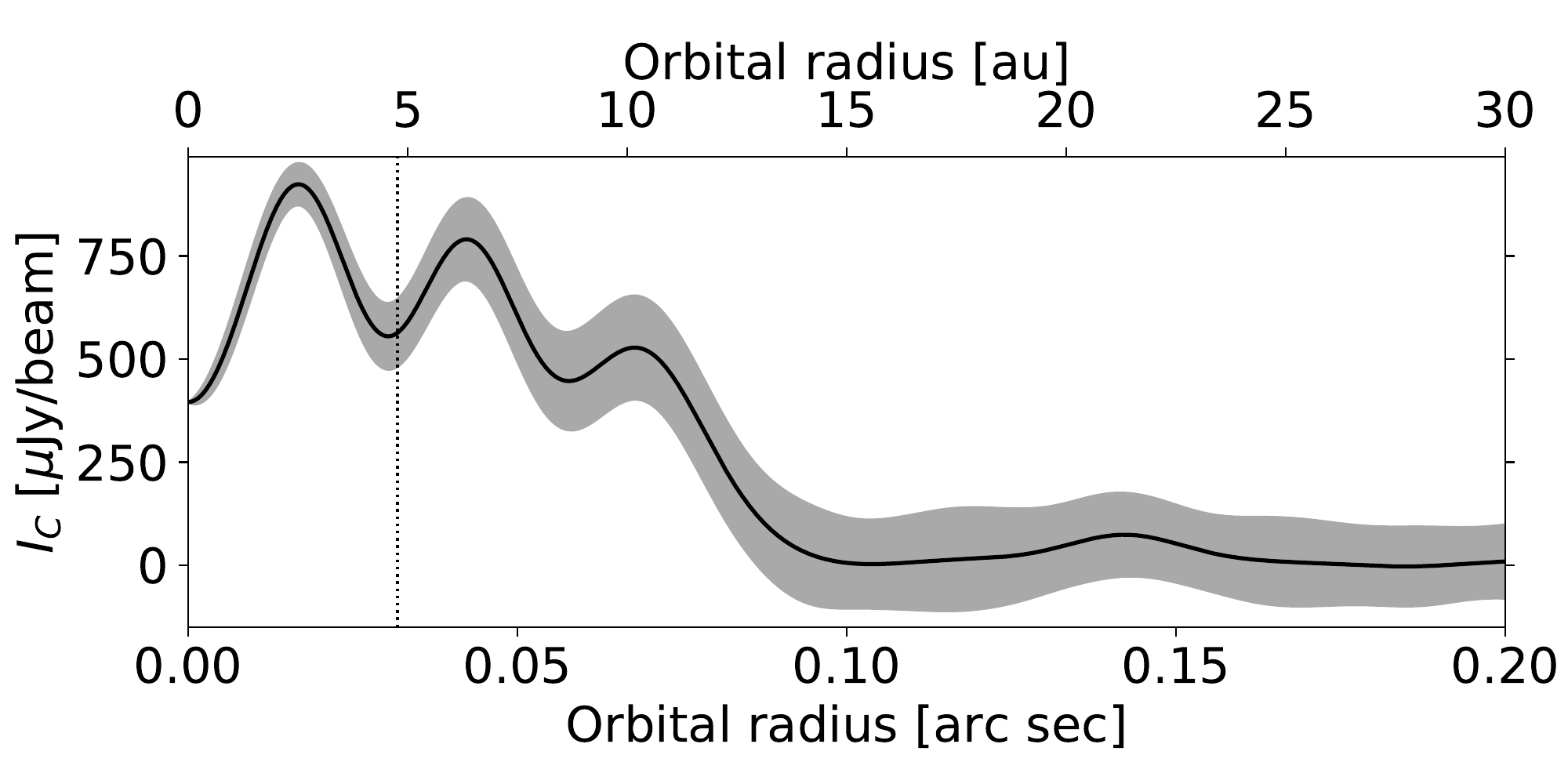}
}
\caption{Same as Fig.~\ref{fig:uter_rt}, but for the Pendragon simulation at 59900 years (669 planet orbits).}
\label{fig:pendragon_rt}
\end{figure}

\subsubsection{Continuum emission at $\lambda=1.3$ mm}
\label{sec:dte_2}
The synthetic map of the 1.3 mm continuum emission obtained from the Pendragon simulation at 59900 years is displayed in the upper panel in Fig.~\ref{fig:pendragon_rt}. The low level of turbulence in this simulation implies sharp structures in the continuum emission. Four bright rings of emission are clearly visible in the synthetic flux map. The two innermost bright rings correspond to the dust rings at the pressure maxima located on both sides of the planet's orbit (which are at around 4 and 7 au, see bottom-right panel of Fig.~\ref{fig:pendragon_case}). The two outermost bright rings correspond to the dust rings that formed outside the planet's orbit prior to each stage of runaway inward migration (they are at about 12 and 24 au, see again the bottom-right panel of Fig.~\ref{fig:pendragon_case}).

We notice that the flux along the bright rings decreases with increasing distance from the star, which mostly reflects the radial decrease in the dust temperature (the emission at the bright rings is optically thick). The intensity of the outermost bright ring at around 24 au is actually quite low (around 100 $\mu$Jy/beam). Thus, when white noise with a rms level of 50 $\mu$Jy/beam is added to the raw flux map, the outermost ring is almost not visible anymore, whereas the innermost rings are still clearly visible (see middle and bottom panels of Fig.~\ref{fig:pendragon_rt}).

\section{discussion and summary}
\label{sec:conclusion}

\subsection{Longevity of the dust rings}
\label{sec:longevity}

\begin{figure}
\centering
\resizebox{1.05\hsize}{!}
{
\hspace{-0.7cm}
\includegraphics[width=1.3\hsize]{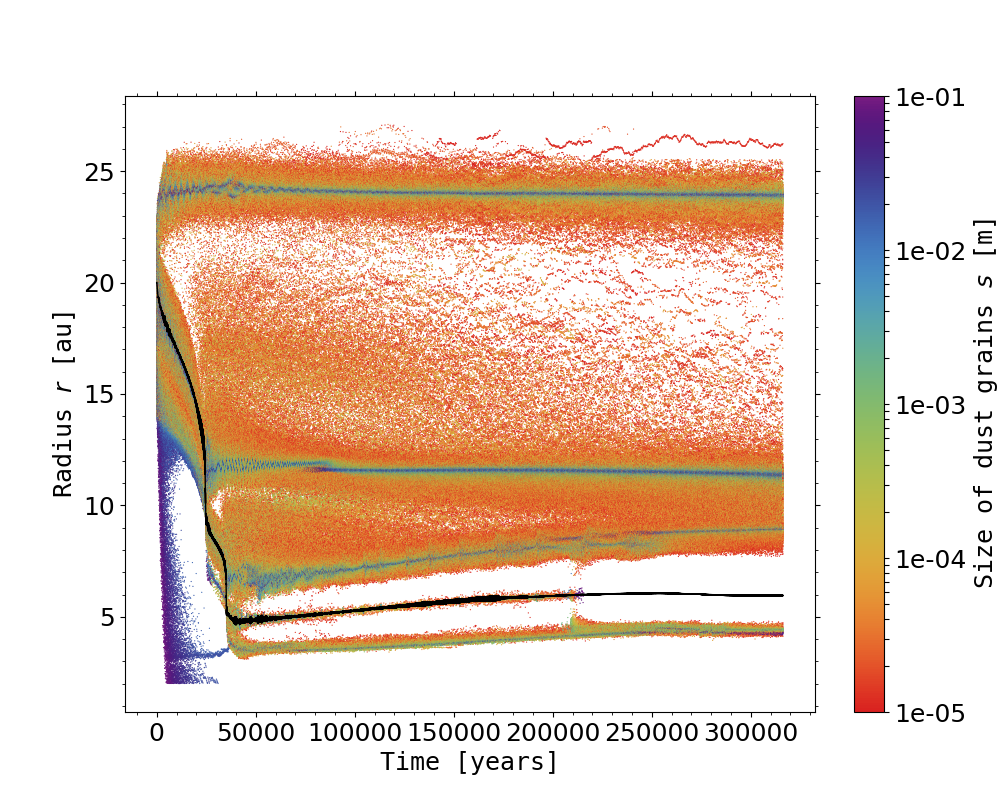}
\hspace{0.7cm}
}
\caption{Same as in Fig.~\ref{fig:spacetime_uther} for an extra run with the same setup as Pendragon but with a turbulent viscosity $\alpha=10^{-5}$.}
\label{fig:spacetime_pendragon_1em5}
\end{figure}
We discuss in this section the longevity of the dust rings induced by intermittent runaway migration. When the planet starts a new episode of runaway migration, the pressure maximum that the planet has formed beyond its orbit is no longer sustained by the planet's outer wake. The pressure maximum is therefore progressively smoothed out by the disk's turbulent viscosity, and so is the dust ring that coincides with the pressure maximum. As we have seen in Figs.~\ref{fig:spacetime_uther} and~\ref{fig:spacetime_pendragon}, the dust rings last between 3000 and 10000 years in the Uther simulation ($\alpha=10^{-3}$), whereas some dust rings survive for more than $10^5$ years
in the Pendragon simulation ($\alpha=10^{-4}$).

This comparison highlights that the lifetime of the dust rings does not simply scale inversely proportional to the gas turbulent viscosity. The main reason is that the lifetime of a dust trap partly depends on how sharp the radial pressure profile is around the pressure maximum, which is related to the duration of the stage of decelerated migration between two successive runaways. This duration is sensitive to the disk's turbulent viscosity, but also to the disk's physical model (e.g., the background gas density profile, see Section~\ref{sec:impact_vort}) and the numerical resolution (see Section~\ref{sec:resolution}).

Finally, Fig.~\ref{fig:spacetime_pendragon_1em5} presents the space-time diagram of the dust's radial location for an additional simulation with a setup identical to that of the Pendragon run, but a turbulent viscosity reduced to $\alpha=10^{-5}$. It shows that all dust rings, except the one coorbital with the planet, last over the entire duration of the simulation ($\gtrsim 3\times 10^5$ years). We notice in this case that the planet slowly migrates outwards after having reached the proximity of the inner wave-killing zone, and so do the dust rings formed inside and outside the planet's orbit after the last episode of runaway migration.

\subsection{Initial gas surface density profile}
\label{sec:impact_vort}
We have seen in Fig.~\ref{fig:general} that increasing the initial surface density $\Sigma_0$ from $3\times10^{-4}$ to $3\times10^{-3}$ radically changes the migration pattern of $\mathcal{P}$, as well as the dust spatial distribution (Fig.~\ref{fig:appendix}). According to Figure~14 of \citet{Masset2003}, which shows the occurrence of runaway migration as a function of disk mass and planet mass, the slow migration case in Fig.~\ref{fig:general} (Section~\ref{sec:overview}) corresponds to non-runaway type I migration, whereas the intermittent and fast cases correspond to type III runaway migration. We now show in this section that, for a given value of $\Sigma_0$, inward runaway migration can also be smooth or discontinuous depending on the slope of the initial gas density profile.

\begin{figure*}
\begin{center}
\includegraphics[width=0.99\hsize]{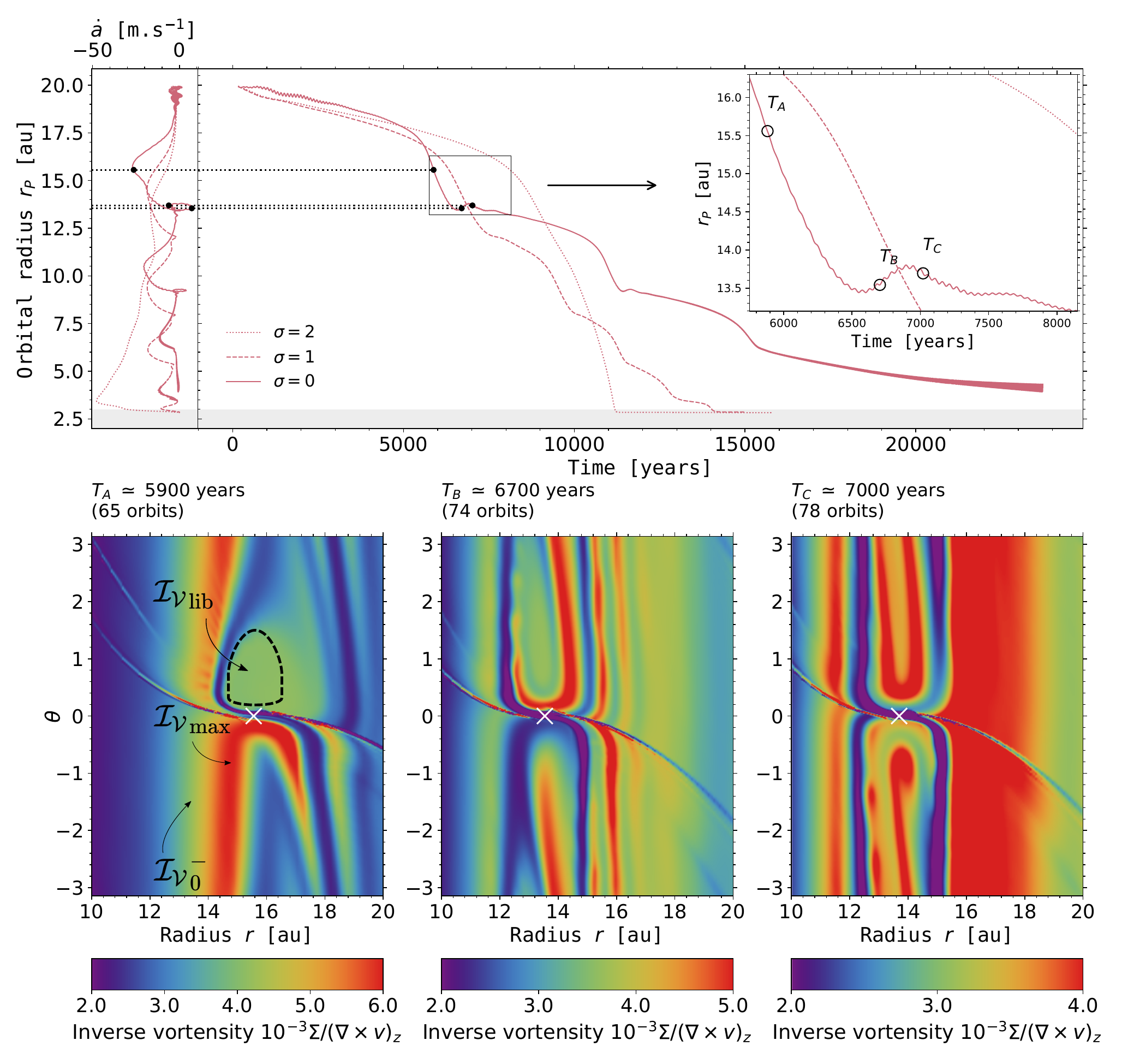}
\end{center}
\caption{Impact of the initial surface density profile of the gas ($\Sigma_0(r) \propto r^{-\sigma}$) on the planet's orbital evolution in the Uther disk model. The upper panel compares the time evolution of the planet's orbital radius and migration rate in the simulation presented in Section~\ref{sec:case1} (dashed curves, $\sigma=1$), with those obtained in two additional simulations for $\sigma = 0$ (solid curves) and $\sigma=2$ (dotted curves). The gray band corresponds to the inner wave-killing zone. The lower panels display three screenshots of the gas inverse vortensity in the $\sigma=0$ simulation at the three times denoted by $T_{\rm A}$, $T_{\rm B}$ and $T_{\rm C}$ in the upper panel. The planet's position is marked by a white cross.}
\label{fig:psig0}
\end{figure*}

By varying the power-law exponent of the initial surface density profile of the gas, that is $\sigma$ in $\Sigma_0(r) = \Sigma_0 \times \left(r/r_0\right)^{-\sigma}$, we change the unperturbed (i.e., initial) inverse vortensity profile, $\mathcal{I_V}_0(r) = \Sigma_0(r)/\omega_0(r) \propto r^{3/2-\sigma}$, and thus the inverse vortensity of the material entering the horseshoe region ($\mathcal{I_V}_e$) as the planet migrates. To examine how different values of $\sigma$ impact our results, we performed two additional simulations with the Uther disk model for $\sigma=0$ and $\sigma=2$, the fiducial Uther simulation presented in Section~\ref{sec:case1} having $\sigma=1$. In the following, we refer to these simulations as the $\sigma_0$, $\sigma_1$ and $\sigma_2$ runs. The upper panel of Fig.~\ref{fig:psig0} compares the time evolution of the planet's orbital radius ($r_{\rm p}$) and migration speed ($\dot{a}$) in these three simulations.

We see that, initially, $|\dot{a}|$ takes very similar values in all three runs ($|\dot{a}| \sim 2$ m s$^{-1}$). From this we argue that the static corotation torque plays a minor role in the initial migration rate. This is confirmed by inspection at Eq. (51) of \citet{Paardekooper2011}, which indicates that the static corotation torque saturates for the values of $q_p$, $h_0$ and $\alpha$ in the Uther simulation. The migration behavior changes dramatically, however, after the first episode of runaway: in the $\sigma_2$ run, migration remains smooth and fast, whereas in the $\sigma_0$ run migration is intermittent, with stages of decelerated migration that last longer than in the $\sigma_1$ run. These behaviors boil down again to the radial profile of inverse vortensity.

In the $\sigma_2$ run, the background inverse vortensity profile decreases with radius (as $r^{-1/2}$). The unperturbed material that crosses the orbit has an inverse vortensity comparable to, if not larger than that of the gas initially shocked by the inner wake. The coorbital vortensity deficit thus does not decrease when the planet starts interacting with the unperturbed gas inside its orbit, and migration does not decelerate. Shock formation is impeded, and runaway migration is fueled solely by the increasing inverse vortensity of the background material that crosses the orbit as $\mathcal{P}$ moves inwards. A similar result was predicted by \citet{Lin2010} (their section 6.1.1). Consequently, migration becomes increasingly fast, and it is clear in the upper panel of Fig.~\ref{fig:psig0} that $|\dot{a}|$ increases nearly continuously in the $\sigma_2$ simulation until $\mathcal{P}$ reaches the grid's inner boundary.

On the contrary, in the $\sigma_0$ and $\sigma_1$ runs, the background inverse vortensity increases with radius. The unperturbed material crossing the orbit has an inverse vortensity smaller than that of the gas initially shocked by the inner wake. The coorbital vortensity deficit therefore decreases when $\mathcal{P}$ starts interacting with this unperturbed material, runaway stalls, and $\mathcal{P}$ is able to generate an $\mathcal{I_V}$ maximum inside its orbit that fuels the next episode of runaway migration.

The maximum migration rate is determined by the difference between the $\mathcal{I_V}$ maximum of the gas shocked by the inner wake (which we denote by $\mathcal{I_V}_{\rm max}$) and $\mathcal{I_V}_{\rm lib}$ (which is close to $\mathcal{I_V}_0(r_{\rm p})$). Both quantities are annotated in the lower-left panel of Fig.~\ref{fig:psig0}. This difference of inverse vortensity increases upon lowering $\sigma$, and we see in the upper-left panel of Fig.~\ref{fig:psig0} that the maximum migration rate increases for smaller $\sigma$, at least in the first episode of runaway migration. Likewise, the braking is determined by the difference between $\mathcal{I_V}_{\rm max}$ and the $\mathcal{I_V}_0$ of the material orbiting inside the shocked gas (denoted by  $\mathcal{I_V}^-_0$ in the lower-left panel of Fig.~\ref{fig:psig0}). Again, this difference increases (in absolute value) for smaller $\sigma$. The braking also depends on the ability of the material entering the horseshoe region to perform secondary U-turns, which slows down migration by adding a positive corotation torque.

We have already mentioned the existence of secondary U-turns for the Uther simulation in Section~\ref{sec:case1} (see the panel at time $T_{\rm D}$ in Fig.~\ref{fig:invortensity}). A smaller $\sigma$ favors multiple U-turns of the material entering the horseshoe region, a larger induced corotation torque, and therefore a more abrupt braking of the migration. A short episode of outward migration is even observed at the end of the first runaway in the $\sigma_0$ run, as highlighted by the inset plot in the upper panel of Fig.~\ref{fig:psig0}. The three panels of inverse vortensity in Fig.~\ref{fig:psig0} help understand this short episode. The left panel at time $T_{\rm A}=5900$ years is when $|\dot{a}|$ is maximum. A trapezoidal region of material trapped in libration with $\mathcal{P}$ forms at $0 \lesssim \theta \lesssim 2$ rad, whereas the inner material with maximum $\mathcal{I_V}$ executes primary U-turns outwards, contributing to the planet's runaway. Part of this material with maximum $\mathcal{I_V}$ crosses the orbit, but when $\mathcal{P}$ starts slowing down, part of it stays in the horseshoe region and performs secondary U-turns. This situation is illustrated in the middle panel at time $T_{\rm B}=6700$ years, where the $\mathcal{I_V}$ difference of the gas just in front of the planet in azimuth (high $\mathcal{I_V}$) and just behind it (low $\mathcal{I_V}$) is large enough for the (positive) corotation torque to allow outward migration. This $\mathcal{I_V}$ difference changes sign in the right panel at time $T_{\rm C}=7000$ years. Progressive advection-diffusion of vortensity in the horseshoe region, and shock formation inside the gap's inner edge, reinstate inward migration. The oscillations in the planet's orbital radius described by \citet{McNally2019} after each runaway phase correspond in our work to secondary U-turns of high-$\mathcal{I_V}$ material in the coorbital region. These secondary U-turns help at once brake the planet's migration and replenish the reservoir of dust particles ahead of the planet before the pending runaway phase.

Concerning the impact of $\sigma$ on the dust, we point out that the number of rings is smaller in the $\sigma_0$ run than in the $\sigma_1$ run, due to the reduced number of episodes of runaway migration. Moreover, because the phases of non-runaway migration last longer for $\sigma_0$, the pressure maxima are sharper, and the intensity ratio between bright and dark rings is larger. We have checked this by a dedicated dust radiative transfer calculation for the $\sigma_0$ run (not shown here). The sharper pressure maxima in the $\sigma_0$ run also imply longer-lived dust rings.

To conclude this section, we stress again that, for a given planet mass and disk's aspect ratio, inward runaway migration comes in two flavors: (i) smooth and fast, or (ii) intermittent and globally slower. Getting one behavior or the other depends on the profile of inverse vortensity in the disk, in particular across the inner edge of the planet gap. As we have seen, a low turbulent viscosity and a shallow gas surface density profile favor intermittent runaway with sharper stairs in the migration pattern.

\subsection{Impact of gas self-gravity}
\label{sec:impact_sg}

\begin{figure}
\centering
\resizebox{1.07\hsize}{!}
{
\hspace{0.15cm}
\includegraphics[width=0.99\hsize]{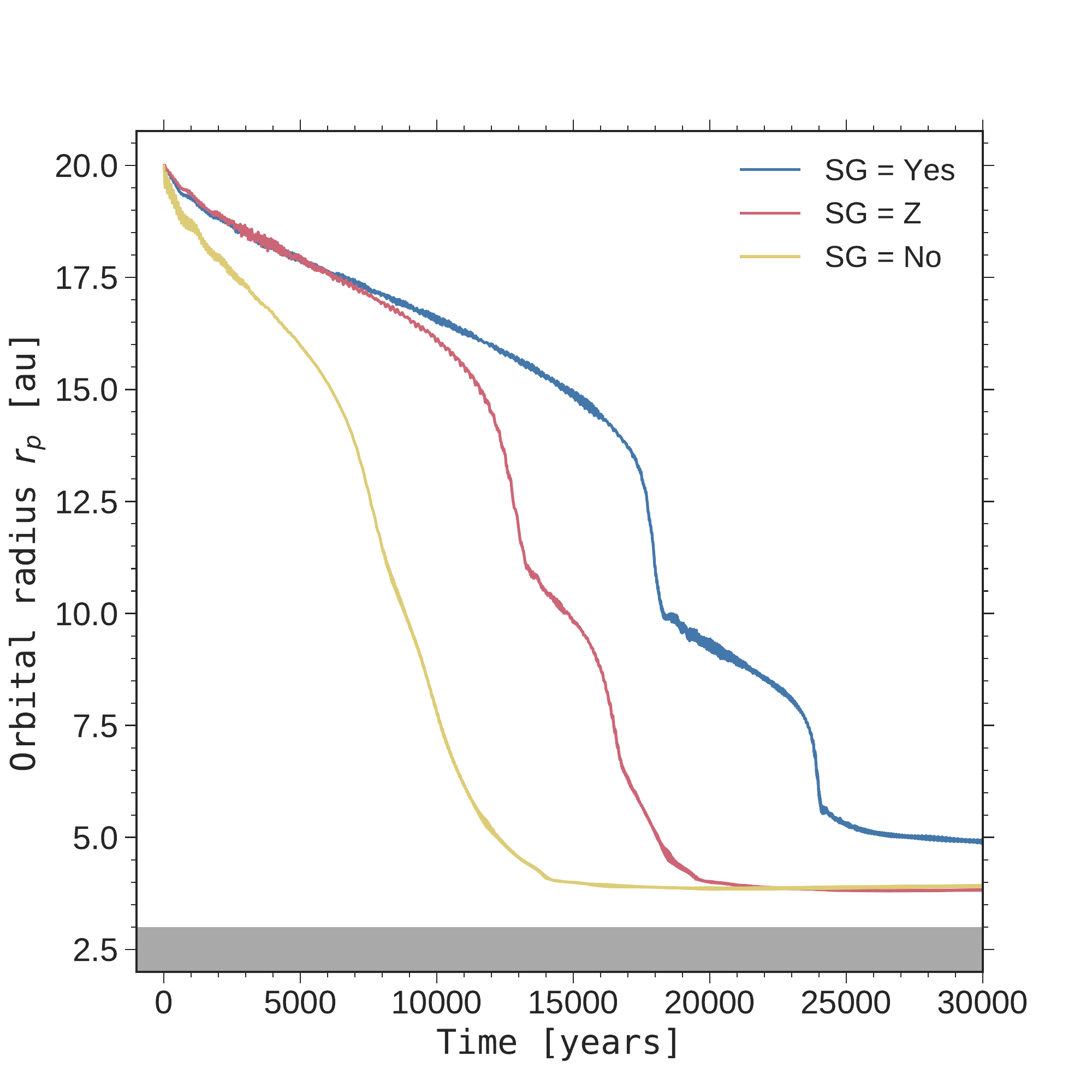}
}
\caption{Impact of including gas self-gravity on the planet's orbital evolution in the Pendragon simulation. Results are shown with full inclusion of the gas self-gravity (blue curve labeled as SG = Yes), with only the axisymmetric part of the gas self-gravity (red curve, SG = Z), and without self-gravity (yellow curve, SG = No). The gray band shows the inner wave-killing zone.}
\label{fig:sg}
\end{figure}

Depending on the disk's aspect ratio $h_0$, the disk models for which intermittent runaway migration is observed in Fig.~\ref{fig:general} have an initial Toomre Q-parameter in the range $[16-19]$ at 10 au, which is about half way between the planet's initial location and the outer edge of the inner damping boundary. These values of Q are large enough to wonder whether gas self-gravity should be included in the simulations. To show that gas self-gravity actually matters despite these rather large Q values, we compare in Fig.~\ref{fig:sg} the planet's orbital evolution in the Pendragon run (blue curve, Section~\ref{sec:case2}) with that in two additional simulations: one that discards gas self-gravity (yellow curve), and another one where only the axisymmetric part of the gas self-gravity is accounted for (red curve, see \citealp{BaruteauMasset2008b}). These two additional simulations use otherwise the same setup as Pendragon.

We see that the initial migration rate is substantially larger in the run without self-gravity. This is due to a spurious inward shift of the Lindblad resonances when the planet feels the gravitational potential of the disk but the disk does not feel its own potential \citep{PH05,BaruteauMasset2008b}. This inward shift causes an increase in the magnitude of the net Lindblad torque that scales with $(Qh)^{-1}$, and which becomes significant when $Qh \lesssim 1$, as is the case in the Pendragon and Uther simulations (see \citealp{BaruteauMasset2008b}, whose results were obtained for non gap-opening planets). As a consequence of the faster (initial) migration, the steps in the migration pattern are much less visible in the run without self-gravity (see Section~\ref{sec:impact_vort}). To understand why this is the case, one has to dwell again on the inverse vortensity profile ($\mathcal{I_V}$). Without self-gravity, we find that the planet is able to maintain a small $\mathcal{I_V}$ bump inside its orbit all over its migration course. The orbit-crossing gas has an $\mathcal{I_V}$ which is high enough to sustain runaway migration, but not high enough that the planet can overcome this $\mathcal{I_V}$ bump, which would cause gas with unperturbed, smaller $\mathcal{I_V}$ to start crossing the orbit, and which would therefore cease runaway, as explained in Section~\ref{sec:cvd}. The smaller amplitude of the $\mathcal{I_V}$ bump inside the planet orbit without self-gravity is due to the artificially faster migration rate in that case, which weakens the shock driven by the planet's inner wake (the synodic period of the gas around the gap's inner edge being too short in comparison to the planet's migration timescale over its horseshoe region). In the run without self-gravity, the planet's migration rate ($|\dot{a}|$) stays around 10 m s$^{-1}$, while with self-gravity $|\dot{a}|$ oscillates between approximately 0 and 20 m s$^{-1}$.

In the simulation with the axisymmetric part of the gas self-gravity, the initial migration rate is very similar to that with full self-gravity, in agreement again with past studies on the Lindblad torque \citep{PH05,BaruteauMasset2008b}. We notice, however, that runaway migration starts earlier in the run with axisymmetric self-gravity. Comparison with full self-gravity points to the presence of stronger (high-$\mathcal{I_V}$) vortices in the the run with axisymmetric self-gravity, in particular along the inner separatrix of the horseshoe region, which fuels the runaway process. This is due to the fact that the non-axisymmetric part of the gas self-gravity significantly damps large-scale vortices induced by the Rossby-Wave Instability (with azimuth wavenumber $m=1$) when $Q\times h \lesssim 1$, as is the case in these simulations \citep[e.g.,][]{lovelace2013,Zhu2016}.

We finally point out that, in some of our simulations with full self-gravity, growth of a global $m=1$ mode is observed (hints of this behavior can be seen in the $\sigma_0$ run displayed in Fig.~\ref{fig:psig0}, through oscillations of increasing amplitude visible in the planet's orbital radius after $\sim 18000$ years). Similar findings have been recently reported \citep{PierensLin2018, Perez2019, Baruteau2019}. The nature and implications of such mode growth require a dedicated study.

\subsection{Inclusion of an energy equation}
\label{sec:impact_nrj}
As already stated in Section~\ref{sec:gas}, a locally isothermal equation of state is more relevant to the outer parts of protoplanetary disks, typically beyond a few tens of au, where the radiative diffusion and/or cooling timescales become shorter than the orbital timescale. This section aims at examining the impact of an energy equation on the results of the Pendragon simulation. The energy equation solves the time evolution of the gas thermal energy density, and it includes viscous heating, radiative cooling and stellar irradiation. However, radiative diffusion is discarded for simplicity. Our energy equation is thus the same as, e.g., Equation 1 of \citet{PierensRaymond16}, but without the radiative diffusion term. The Rossland mean opacity is taken from \citet{BellLin94}. The gas adiabatic index is set to $5/3$, and for our fiducial profiles of initial surface density and temperature, the initial entropy profile happens to be nearly flat ($\propto r^{-1/30}$). Since viscous heating is small in the Pendragon disk model ($\alpha = 10^{-4}$), the gas temperature very quickly evolves towards the temperature profile set by stellar irradiation. For comparison purposes, the radial profile of the irradiation temperature ($T_{\rm irr}$) is chosen such that the radial profile of the gas adiabatic sound speed approximately matches that of the isothermal sound speed in the Pendragon run, which led us to adopt $T_{\rm irr}(r) = 80\,{\rm K}\times (r/r_0)^{-0.7}$ (recall that $r_0 = 10$ au). In the following, simulations that include the above energy equation are referred to as radiative simulations, those that do not as isothermal simulations.

\begin{figure}
\centering
\resizebox{1.01\hsize}{!}
{
\hspace{0.35cm}
\includegraphics[width=0.99\hsize]{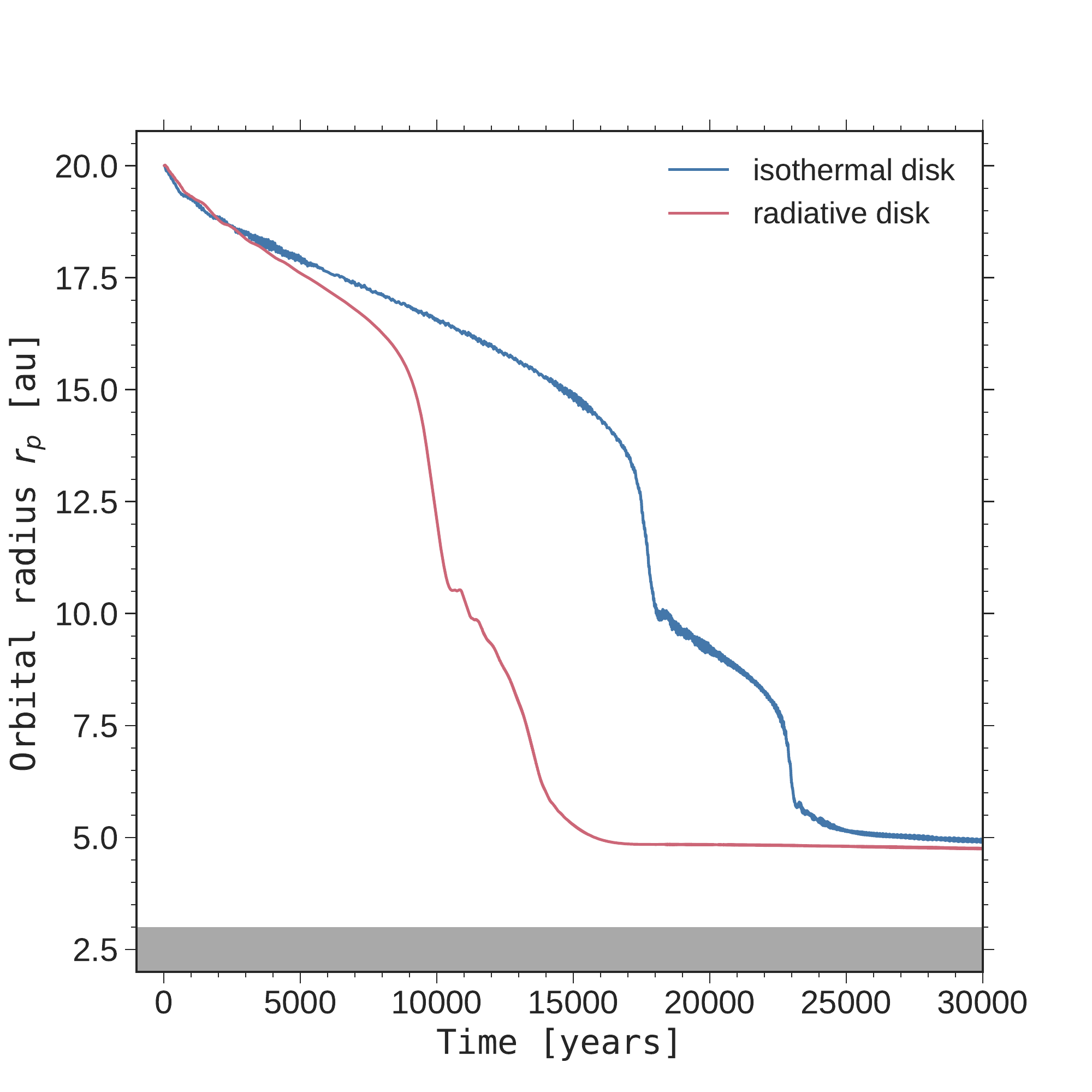}
}
\resizebox{0.94\hsize}{!}
{
\includegraphics[width=0.99\hsize]{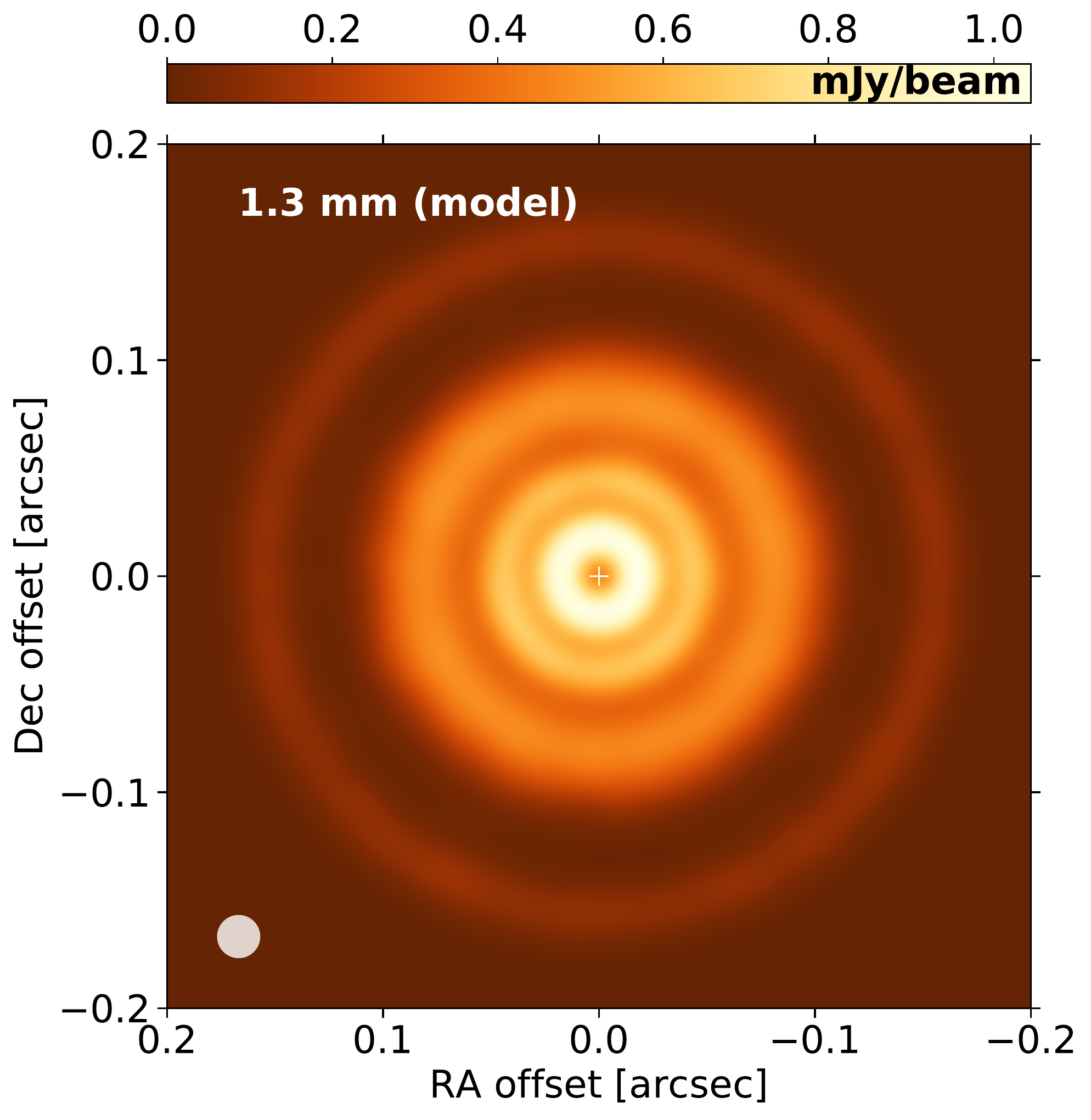}
}
\resizebox{0.92\hsize}{!}
{
\hspace{0.2cm}
\includegraphics[width=0.99\hsize]{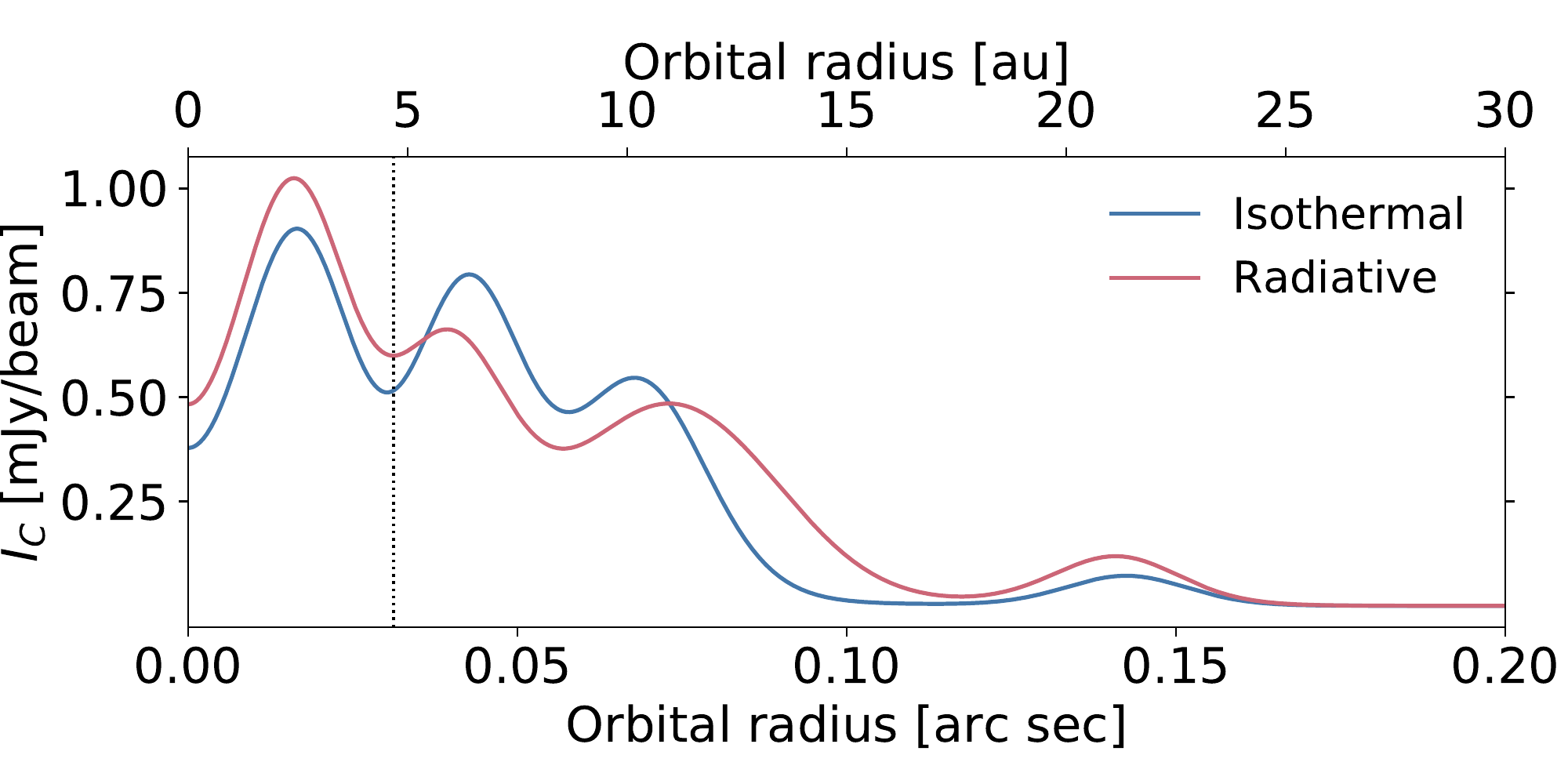}
}
\caption{\textbf{Top:} Time evolution of the planet's orbital radius in the Pendragon run (blue curve, labeled as isothermal disk) and in an additional run that includes an energy equation (red curve, labeled as radiative disk; see text). The gray band shows the inner wave-killing zone. \textbf{Middle:} Synthetic flux map of the continuum emission at $\lambda = 1.3$ mm in the radiative run, convolved with a $0\farcs02$ circular beam, at $\sim 40000$ years ($\sim 25000$ years after migration halted). \textbf{Bottom:} Azimuthally-averaged convolved intensity versus distance to the star in the isothermal and radiative runs. The intensity of the isothermal run is at 59900 years (also $\sim 25000$ years after migration stopped in that run). The vertical dotted line shows the planet location.}
\label{fig:nrj}
\end{figure}
In the upper panel of Fig.~\ref{fig:nrj}, we compare the time evolution of the planet's orbital radius in the isothermal and radiative Pendragon simulations. The migration pattern of $\mathcal{P}$ is similar in both simulations, with basically two stages of runaway migration that cease at about the same locations in the disk (near 10 and 5 au). The initial migration rate is somewhat higher in the radiative run, however, and the first runaway occurs earlier. The origin of this behavior is not entirely clear. One possible lead is the absence of vortices observed in the radiative run, or the fact the planetary shocks appear to be smoother in that run. A detailed investigation would require a specific study. In agreement with the results presented in Section~\ref{sec:impact_vort}, this overall faster migration results in smoother stairs. Nonetheless, the similar pattern of intermittent migration leads to the formation of similar dust structures at the end of the radiative simulation, and therefore a similar sequence of bright and dark rings of continuum emission, as illustrated in the middle and bottom panels in Fig.~\ref{fig:nrj}. Because the stage of decelerated migration at $\sim$10 au lasts shorter in the radiative simulation, the dust ring that forms at the pressure maximum outside the planet (near 12 au) fades away faster than in the isothermal simulation. The lower panel in Fig.~\ref{fig:nrj} actually makes it clear that the bright ring of continuum emission near 12 au is broader in the radiative simulation.

For the sake of completeness, we also performed a radiative simulation with the Uther disk model ($\alpha = 10^{-3}$). The profile of irradiation temperature was chosen as $T_{\rm irr}(r) = 60\,{\rm K}\times (r/r_0)^{-0.7}$. Although not shown here, the migration pattern in this radiative simulation is very smooth, and the absence of clear steps in the migration, along with the higher level of disk turbulence, preclude the formation of well-defined dust rings in the disk. Still, we have checked that intermittent migration and the formation of multiple dust rings are recovered in the Uther radiative simulation by simply increasing the planet mass.

\subsection{Grid resolution}
\label{sec:resolution}

\begin{figure}
\centering
\resizebox{1.07\hsize}{!}
{
\hspace{0.15cm}
\includegraphics[width=0.99\hsize]{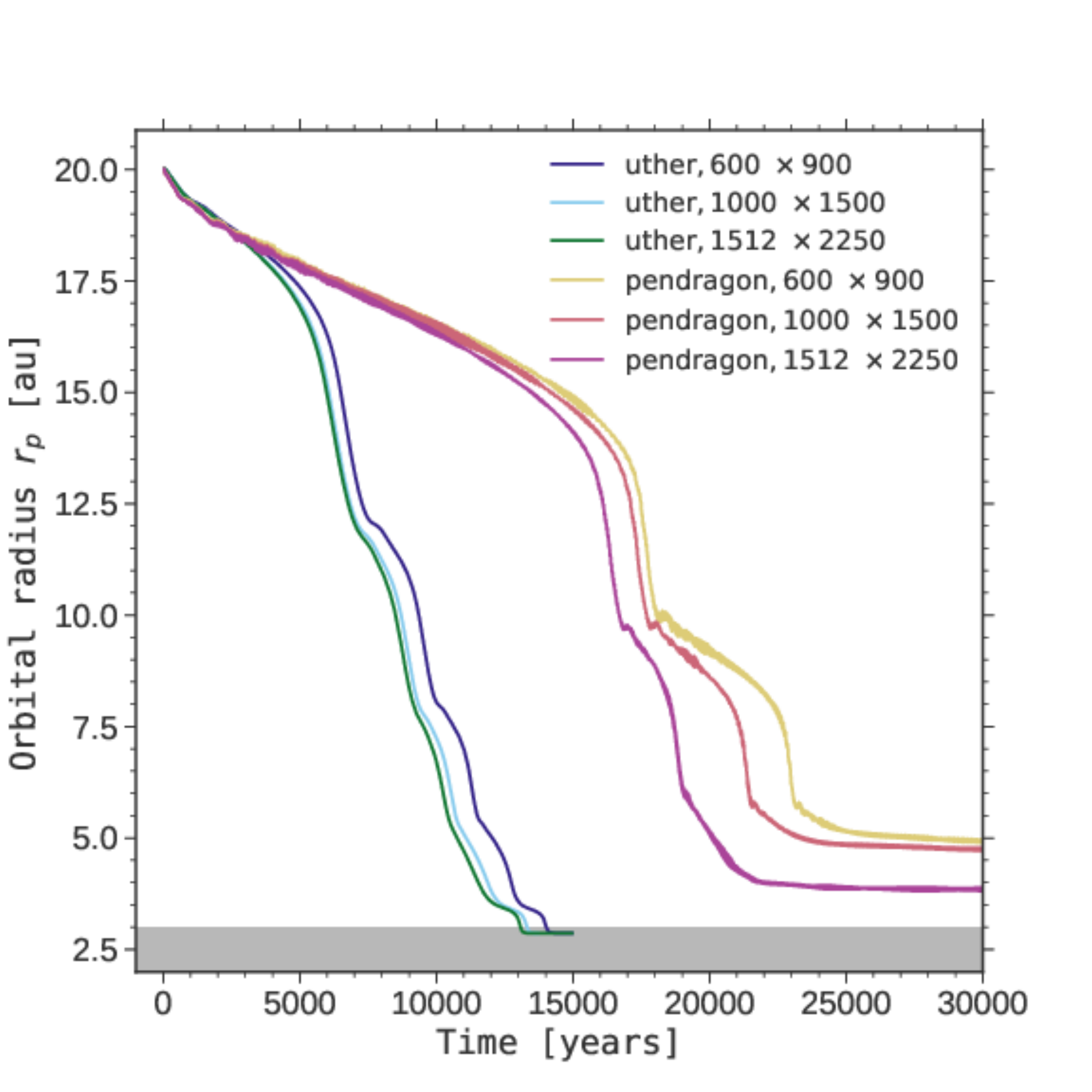}
}
\caption{Impact of increasing grid resolution on the planet's orbital evolution in the Uther and Pendragon simulations. Results are shown for three grid resolutions : $600\times900$ (fiducial), $1000\times1500$ and $1512\times2250$. The gray band shows the inner wave-killing zone.}
\label{fig:reso}
\end{figure}

When the grid resolution is increased, we find that the planet migrates globally faster, and that, as we have seen in Section~\ref{sec:impact_vort}, the stages of decelerated migration last shorter. This is illustrated in Fig.~\ref{fig:reso} for the Uther and Pendragon simulations, where the fiducial grid resolution is increased to $1000\times1500$ and $1512\times2250$. The convergence in resolution is more easily achieved in the Uther run due to the larger disk's turbulent viscosity. Increasing resolution allows to better resolve small-scale structures in the gas vortensity within or near the planet's horseshoe region. This is particularly critical to the simulations at low viscosity, like in the Pendragon run with $\alpha=10^{-4}$, for which increasing resolution tends to form smaller, high-$\mathcal{I_V}$ vortices along the inner separatrix of the horseshoe region, which therefore embark more easily onto U-turns, and boost the planet's migration rate. These small-scale vortices, which are better resolved in the $1512\times2250$ simulation, thus tend to extend the runaway phase, and bring the planet closer to the grid's inner boundary. Similar results have been obtained by \citet{McNally2019}, who examined intermittent runaway migration in low-viscosity and inviscid disk models, and found that convergence in resolution was increasingly difficult to achieve, if achievable at all, upon decreasing viscosity.

Although, in our simulations, stages of decelerated migration last shorter when increasing resolution, a pressure maximum does form when migration decelerates, but it is smoother, the density of dust trapped at this location is therefore smaller, and so is the intensity of the corresponding bright ring. For Pendragon, for instance, when the number of grid cells is increased from $600\times900$ to $1000\times1500$, the intensity of the 3 innermost bright rings is decreased by a few percent up to 30$\%$, depending on the ring. Concerning the longevity of dust rings, for example the ring at 12 au in the Pendragon simulation, its lifetime is shorter when increasing resolution because the preceding stage of slow migration is shorter and the corresponding pressure maximum therefore smoother. For the $1512\times2250$ run, the ring's disruption starts at $\sim$50000 years, while it does so at $\sim$70000 years at $600\times900$.

\subsection{Concluding remarks}
\label{sec:ccl}

We have shown in this study that a single planet undergoing multiple episodes of runaway inward migration could form several dust rings and gaps in its protoplanetary disk, and therefore sequences of bright and dark rings of continuum emission at (sub)mm wavelengths. We have seen that the gas inverse vortensity is the key quantity that grasps the planet's orbital evolution all over the stages of runaway migration, and that the formation of the dust rings boils down to the creation of a pressure maximum or an inflection point on both sides of the planet's orbit every time runaway migration ceases. As runaway stalls, the gas that enters the planet's horseshoe region exerts a positive corotation torque through secondary U-turns, which further decelerates the planet migration. When the planet decelerates, its wakes have enough time to shock the gas and thus rebuild a pressure maximum or an inflection point on both sides of the planet's orbit, which efficiently trap the largest dust particles. It is the trapping of these large particles at the pressure maximum outside the planet's partial gap which causes a new bright ring in the continuum emission at radio wavelengths. By lengthening the stages of non-runaway migration, the aforementioned secondary U-turns help enhance the pressure maximum outside the planet gap, which increases the density contrast between the dust gap and the dust ring, and therefore the intensity contrast between the corresponding dark and bright rings of emission. A longer stage of non-runaway migration also helps increase the longevity of the dust ring that forms at the pressure maximum.

The {\it N Gaps N Planets} (NGNP) model mentioned in Section~\ref{sec:intro} needs as many fixed planets as observed dark rings in the disk's continuum emission. However, with our scenario of intermittent runaway migration, planet-induced bright rings can be obtained without expecting planets in the vicinity of gaps in the continuum emission. As already stated by \citet{McNally2019}, an intermittently migrating planet can wind up close to the star and be unresolved for the current observing facilities, and still have given rise to lasting multiple rings, depending on the disk's turbulent viscosity.

It has been shown that a single planet on a fixed circular orbit could form more than three bright rings and two dark rings of continuum emission in low-viscosity disks, due to the formation of a tertiary gap inside the planet's orbit \citep{Bae2017multiple, Dong2017multiple, Zhang2018}. We do not find this behavior with the regime of parameters that we have adopted in this study, whether the planet migrates slowly (see Appendix~\ref{sec:appendix_a}) or even remains on a fixed circular orbit. Although numerical artefacts due to the inner damping region play a role, we think that it is likely due to the values of $\alpha$ that we consider in this work, which are too high to allow for a tertiary gap to form.

Our work extends previous studies on intermittent runaway migration by including the full calculation of the gas self-gravity, by testing the effects of an energy equation for the gas, by considering a broad range of sizes for dust particles, and by analyzing the impact of intermittent migration on the disk's continuum emission through dust radiative transfer calculations. In agreement with previous works, we find that runaway migration can be either smooth or discontinuous (intermittent). We actually find a wide range of disk parameters for which runaway migration is intermittent. Intermittence is favored for smaller $\alpha$ turbulent viscosities, flatter gas surface density profiles (smaller $\sigma$, see Section~\ref{sec:impact_vort}), which is globally in line with disks' observational constraints. Although not shown here, we find that intermittence is also favored for larger planet-to-primary mass ratios $q_{\rm p}$ (but still in the partial gap opening regime). We have checked the latter dependence for the Uther disk model presented in Section~\ref{sec:case1} by increasing the planet mass from $0.9$ to $2.7$ Saturn masses. For all these reasons, we believe that the intermittent runaway migration of a giant planet is a viable mechanism to account for the multiple rings and gaps observed in massive protoplanetary disks. Still, rapid planet growth beyond the mass range of intermittent runaway migration may occur, which needs to be examined to further assess the robustness of our mechanism.

Among the possible extensions of this work, it would be interesting to examine how the formation of multiple dust rings and gaps due to intermittent runaway migration would look like in scattered light at near-IR wavelengths. This question would probably be best addressed by dust radiative transfer calculations based on 3D gas and dust hydrodynamical simulations (instead of 2D, as presented in this work). It is possible that intermittent migration may leave imprints in scattered light as radial minima and maxima in the polarized intensity, which would reflect more or less the radial variations of the gas density in the superficial layers of the disk. It would also be interesting to examine the impact of additional planets on our results. Furthermore, we have discarded dust drag onto the gas in our simulations, which we have justified {\it a posteriori} by showing that the dust-to-gas density ratio remains smaller than unity for the dust mass and size distribution adopted in our dust radiative transfer calculations. A larger dust mass or a shallower size distribution could well lead to larger dust-to-gas density ratios, and future work on our migration scenario should therefore aim at including the back-reaction of the dust onto the gas.

Finally, we have seen in the Uther simulation that the dust rings have a geometric spacing $\delta=r_{i+1}/r_{i}\sim 1.5$, with $r_{i}$ the radial location of the $i^{\rm th}$ bright ring. Additional simulations indicate that, when intermittent runaway migration is at work, this geometric spacing mainly depends on the disk's aspect ratio $h_0$, but hardly on $q_{\rm p}$, $\alpha$ \citep[Fig. 11 in][]{Lin2010}, the initial location of the planet $r_{\rm p,0}$, $\Sigma_0$ \citep[Fig. 18 in][]{Lin2010}, and even $\sigma$. A simple estimate of $\delta$ can be obtained by assuming (i) an ideal staircase migration pattern, and (ii) that the radial location of the $\mathcal{I_V}$ maximum outside the planet just after runaway migration coincides with that of the $\mathcal{I_V}$ maximum inside the planet just before runaway migration (Section~\ref{sec:cvd}). This leads to $\delta=(r_{\rm p}+\epsilon)/(r_{\rm p}-\epsilon)$, with $\epsilon$ the radial separation between the planet and the $\mathcal{I_V}$ maxima. The precise value of $\epsilon$ is uncertain, but roughly $\epsilon \gtrsim x_{\rm s}$ (recall that $x_{\rm s}$ denotes the radial half-width of the planet's horseshoe region). We thus expect $\epsilon \propto r_{\rm p}$, and therefore $\delta$ independent of $r_{\rm p}$, which accounts for the geometric spacing of the dust rings. By adopting $\epsilon = x_{\rm s}$, we obtain $\delta \approx 1.25$. This is smaller than the typical value $\sim 1.5$ found in our simulations, which is due in part to the fact that the migration pattern is not an ideal staircase (due to the stages of non-runaway migration in-between the runaway phases). Interestingly, some protoplanetary disks display a very similar geometric spacing for the radial location of successive bright rings in the continuum emission. For RU Lup, we infer $\delta \sim 1.4$ by considering the four bright rings reported in \citet{Huang2018}, and for HD 163296, we infer $\delta \sim 1.5$ by selecting the three outermost rings in \citet{Huang2018}. This spacing could be an interesting quantity to look at in future works to discriminate between different models of multiple rings formation, for a given disk observation. Another quantity worth looking at is the deviation of the gas azimuthal velocity from Keplerian rotation \citep{Teague2018,Perez2018}. It would be interesting to see whether measuring this deviation across the multiple gaps could help differentiate between observations of multiple gaps formed by a single slowly migrating planet, a single planet experiencing intermittent runaway migration (as in this paper), and by multiple planets (NGNP model).

\section*{Acknowledgments}
We acknowledge Florian Thuillet's contribution to the early stages of this work, as part of his Master's thesis (University of Toulouse, 2016). We thank Quentin Kriaa for providing Python scripts that were used to produce some of the figures in this paper. We thank Olivier Bern\'e, Andr\'es Carmona and Fr\'ed\'eric Masset for stimulating discussions. We extend special thanks to Fr\'ed\'eric Masset for a thorough reading of a first draft of this manuscript. We also thank the referee for a thorough review and constructive comments. Numerical simulations were performed on the CALMIP Supercomputing Center of the University of Toulouse.

\bibliographystyle{mnras}

\begin{thebibliography}{}
\makeatletter
\relax
\def\mn@urlcharsother{\let\do\@makeother \do\$\do\&\do\#\do\^\do\_\do\%\do\~}
\def\mn@doi{\begingroup\mn@urlcharsother \@ifnextchar [ {\mn@doi@}
  {\mn@doi@[]}}
\def\mn@doi@[#1]#2{\def\@tempa{#1}\ifx\@tempa\@empty \href
  {http://dx.doi.org/#2} {doi:#2}\else \href {http://dx.doi.org/#2} {#1}\fi
  \endgroup}
\def\mn@eprint#1#2{\mn@eprint@#1:#2::\@nil}
\def\mn@eprint@arXiv#1{\href {http://arxiv.org/abs/#1} {{\tt arXiv:#1}}}
\def\mn@eprint@dblp#1{\href {http://dblp.uni-trier.de/rec/bibtex/#1.xml}
  {dblp:#1}}
\def\mn@eprint@#1:#2:#3:#4\@nil{\def\@tempa {#1}\def\@tempb {#2}\def\@tempc
  {#3}\ifx \@tempc \@empty \let \@tempc \@tempb \let \@tempb \@tempa \fi \ifx
  \@tempb \@empty \def\@tempb {arXiv}\fi \@ifundefined
  {mn@eprint@\@tempb}{\@tempb:\@tempc}{\expandafter \expandafter \csname
  mn@eprint@\@tempb\endcsname \expandafter{\@tempc}}}

\bibitem[\protect\citeauthoryear{{Andrews}}{{Andrews}}{2015}]{Andrews15}
{Andrews} S.~M.,  2015, \mn@doi [\pasp] {10.1086/683178}, \href
  {https://ui.adsabs.harvard.edu/abs/2015PASP..127..961A} {127, 961}

\bibitem[\protect\citeauthoryear{{Ataiee}, {Baruteau}, {Alibert}  \&
  {Benz}}{{Ataiee} et~al.}{2018}]{Ataiee18}
{Ataiee} S.,  {Baruteau} C.,  {Alibert} Y.,   {Benz} W.,  2018, \mn@doi [\aap]
  {10.1051/0004-6361/201732026}, \href
  {https://ui.adsabs.harvard.edu/abs/2018A%26A...615A.110A} {615, A110}

\bibitem[\protect\citeauthoryear{{Avenhaus} et~al.,}{{Avenhaus}
  et~al.}{2018}]{Avenhaus2018}
{Avenhaus} H.,  et~al., 2018, \mn@doi [\apj] {10.3847/1538-4357/aab846}, \href
  {https://ui.adsabs.harvard.edu/abs/2018ApJ...863...44A} {863, 44}

\bibitem[\protect\citeauthoryear{{Bae}, {Zhu}  \& {Hartmann}}{{Bae}
  et~al.}{2016}]{Bae2016}
{Bae} J.,  {Zhu} Z.,   {Hartmann} L.,  2016, \mn@doi [\apj]
  {10.3847/0004-637X/819/2/134}, \href
  {http://adsabs.harvard.edu/abs/2016ApJ...819..134B} {819, 134}

\bibitem[\protect\citeauthoryear{{Bae}, {Zhu}  \& {Hartmann}}{{Bae}
  et~al.}{2017}]{Bae2017multiple}
{Bae} J.,  {Zhu} Z.,   {Hartmann} L.,  2017, \mn@doi [\apj]
  {10.3847/1538-4357/aa9705}, \href
  {https://ui-adsabs-harvard-edu.insu.bib.cnrs.fr/abs/2017ApJ...850..201B}
  {850, 201}

\bibitem[\protect\citeauthoryear{{Barge} \& {Sommeria}}{{Barge} \&
  {Sommeria}}{1995}]{BargeSommeria1995}
{Barge} P.,  {Sommeria} J.,  1995, \aap, \href
  {http://cdsads.u-strasbg.fr/abs/1995A%26A...295L...1B} {295, L1}

\bibitem[\protect\citeauthoryear{{Baruteau} \& {Masset}}{{Baruteau} \&
  {Masset}}{2008a}]{BaruteauMasset2008a}
{Baruteau} C.,  {Masset} F.,  2008a, \mn@doi [\apj] {10.1086/523667}, \href
  {http://adsabs.harvard.edu/abs/2008ApJ...672.1054B} {672, 1054}

\bibitem[\protect\citeauthoryear{{Baruteau} \& {Masset}}{{Baruteau} \&
  {Masset}}{2008b}]{BaruteauMasset2008b}
{Baruteau} C.,  {Masset} F.,  2008b, \mn@doi [\apj] {10.1086/529487}, \href
  {http://adsabs.harvard.edu/abs/2008ApJ...678..483B} {678, 483}

\bibitem[\protect\citeauthoryear{{Baruteau} \& {Zhu}}{{Baruteau} \&
  {Zhu}}{2016}]{Baruteau2016}
{Baruteau} C.,  {Zhu} Z.,  2016, \mn@doi [\mnras] {10.1093/mnras/stv2527},
  \href {http://adsabs.harvard.edu/abs/2016MNRAS.458.3927B} {458, 3927}

\bibitem[\protect\citeauthoryear{{Baruteau} et~al.,}{{Baruteau}
  et~al.}{2014}]{BaruteauPP6}
{Baruteau} C.,  et~al., 2014, \mn@doi [Protostars and Planets VI]
  {10.2458/azu_uapress_9780816531240-ch029}, \href
  {http://adsabs.harvard.edu/abs/2014prpl.conf..667B} {pp 667--689}

\bibitem[\protect\citeauthoryear{{Baruteau} et~al.,}{{Baruteau}
  et~al.}{2019}]{Baruteau2019}
{Baruteau} C.,  et~al., 2019, \mn@doi [\mnras] {10.1093/mnras/stz802}, \href
  {https://ui.adsabs.harvard.edu/abs/2019MNRAS.486..304B} {486, 304}

\bibitem[\protect\citeauthoryear{{Bell} \& {Lin}}{{Bell} \&
  {Lin}}{1994}]{BellLin94}
{Bell} K.~R.,  {Lin} D.~N.~C.,  1994, \mn@doi [\apj] {10.1086/174206}, \href
  {https://ui.adsabs.harvard.edu/abs/1994ApJ...427..987B} {427, 987}

\bibitem[\protect\citeauthoryear{{Ben{\'\i}tez-Llambay}, {Ramos}, {Beaug{\'e}}
  \& {Masset}}{{Ben{\'\i}tez-Llambay} et~al.}{2016}]{Benitez-Llambay2016}
{Ben{\'\i}tez-Llambay} P.,  {Ramos} X.~S.,  {Beaug{\'e}} C.,   {Masset} F.~S.,
  2016, \mn@doi [\apj] {10.3847/0004-637X/826/1/13}, \href
  {https://ui.adsabs.harvard.edu/abs/2016ApJ...826...13B} {826, 13}

\bibitem[\protect\citeauthoryear{{B{\'e}thune}, {Lesur}  \&
  {Ferreira}}{{B{\'e}thune} et~al.}{2017}]{Bethune2017}
{B{\'e}thune} W.,  {Lesur} G.,   {Ferreira} J.,  2017, \mn@doi [\aap]
  {10.1051/0004-6361/201630056}, \href
  {http://cdsads.u-strasbg.fr/abs/2017A%26A...600A..75B} {600, A75}

\bibitem[\protect\citeauthoryear{{Birnstiel}, {Klahr}  \&
  {Ercolano}}{{Birnstiel} et~al.}{2012}]{Birnstiel2012}
{Birnstiel} T.,  {Klahr} H.,   {Ercolano} B.,  2012, \mn@doi [\aap]
  {10.1051/0004-6361/201118136}, \href
  {https://ui.adsabs.harvard.edu/abs/2012A&A...539A.148B} {539, A148}

\bibitem[\protect\citeauthoryear{{Casoli} \& {Masset}}{{Casoli} \&
  {Masset}}{2009}]{Casoli2009}
{Casoli} J.,  {Masset} F.~S.,  2009, \mn@doi [\apj]
  {10.1088/0004-637X/703/1/845}, \href
  {https://ui.adsabs.harvard.edu/abs/2009ApJ...703..845C} {703, 845}

\bibitem[\protect\citeauthoryear{{Dipierro} \& {Laibe}}{{Dipierro} \&
  {Laibe}}{2017}]{Dipierro17}
{Dipierro} G.,  {Laibe} G.,  2017, \mn@doi [\mnras] {10.1093/mnras/stx977},
  \href {https://ui.adsabs.harvard.edu/abs/2017MNRAS.469.1932D} {469, 1932}

\bibitem[\protect\citeauthoryear{{Dipierro}, {Price}, {Laibe}, {Hirsh},
  {Cerioli}  \& {Lodato}}{{Dipierro} et~al.}{2015}]{Dipierro2015}
{Dipierro} G.,  {Price} D.,  {Laibe} G.,  {Hirsh} K.,  {Cerioli} A.,   {Lodato}
  G.,  2015, \mn@doi [\mnras] {10.1093/mnrasl/slv105}, \href
  {https://ui.adsabs.harvard.edu/abs/2015MNRAS.453L..73D} {453, L73}

\bibitem[\protect\citeauthoryear{{Dong}, {Li}, {Chiang}  \& {Li}}{{Dong}
  et~al.}{2017}]{Dong2017multiple}
{Dong} R.,  {Li} S.,  {Chiang} E.,   {Li} H.,  2017, \mn@doi [\apj]
  {10.3847/1538-4357/aa72f2}, \href
  {https://ui-adsabs-harvard-edu.insu.bib.cnrs.fr/abs/2017ApJ...843..127D}
  {843, 127}

\bibitem[\protect\citeauthoryear{{Draine} \& {Lee}}{{Draine} \&
  {Lee}}{1984}]{Draine1984}
{Draine} B.~T.,  {Lee} H.~M.,  1984, \mn@doi [\apj] {10.1086/162480}, \href
  {http://adsabs.harvard.edu/abs/1984ApJ...285...89D} {285, 89}

\bibitem[\protect\citeauthoryear{{Dullemond}, {Juhasz}, {Pohl}, {Sereshti},
  {Shetty}, {Commercon}  \& {Flock}}{{Dullemond} et~al.}{2015}]{Dullemond2015}
{Dullemond} C.,  {Juhasz} A.,  {Pohl} A.,  {Sereshti} F.,  {Shetty}
  R.and~{Peters} T.,  {Commercon} B.,   {Flock} M.,  2015, {RADMC3D},
  \url{http://www.ita.uni-heidelberg.de/\~dullemond/software/radmc-3d/}

\bibitem[\protect\citeauthoryear{{Fedele} et~al.,}{{Fedele}
  et~al.}{2017}]{Fedele2017}
{Fedele} D.,  et~al., 2017, \mn@doi [\aap] {10.1051/0004-6361/201629860}, \href
  {https://ui.adsabs.harvard.edu/abs/2017A&A...600A..72F} {600, A72}

\bibitem[\protect\citeauthoryear{{Flaherty}, {Hughes}, {Rosenfeld}, {Andrews},
  {Chiang}, {Simon}, {Kerzner}  \& {Wilner}}{{Flaherty}
  et~al.}{2015}]{Flaherty2015}
{Flaherty} K.~M.,  {Hughes} A.~M.,  {Rosenfeld} K.~A.,  {Andrews} S.~M.,
  {Chiang} E.,  {Simon} J.~B.,  {Kerzner} S.,   {Wilner} D.~J.,  2015, \mn@doi
  [\apj] {10.1088/0004-637X/813/2/99}, \href
  {http://cdsads.u-strasbg.fr/abs/2015ApJ...813...99F} {813, 99}

\bibitem[\protect\citeauthoryear{{Flock}, {Ruge}, {Dzyurkevich}, {Henning},
  {Klahr}  \& {Wolf}}{{Flock} et~al.}{2015}]{Flock15}
{Flock} M.,  {Ruge} J.~P.,  {Dzyurkevich} N.,  {Henning} T.,  {Klahr} H.,
  {Wolf} S.,  2015, \mn@doi [\aap] {10.1051/0004-6361/201424693}, \href
  {http://adsabs.harvard.edu/abs/2015A%26A...574A..68F} {574, A68}

\bibitem[\protect\citeauthoryear{{Fuente} et~al.,}{{Fuente}
  et~al.}{2017}]{Fuente2017}
{Fuente} A.,  et~al., 2017, \mn@doi [\apjl] {10.3847/2041-8213/aa8558}, \href
  {http://adsabs.harvard.edu/abs/2017ApJ...846L...3F} {846, L3}

\bibitem[\protect\citeauthoryear{{Goodman} \& {Rafikov}}{{Goodman} \&
  {Rafikov}}{2001}]{Goodman2001}
{Goodman} J.,  {Rafikov} R.~R.,  2001, \mn@doi [\apj] {10.1086/320572}, \href
  {https://ui.adsabs.harvard.edu/abs/2001ApJ...552..793G} {552, 793}

\bibitem[\protect\citeauthoryear{{Huang} et~al.,}{{Huang}
  et~al.}{2018}]{Huang2018}
{Huang} J.,  et~al., 2018, \mn@doi [\apjl] {10.3847/2041-8213/aaf740}, \href
  {http://cdsads.u-strasbg.fr/abs/2018ApJ...869L..42H} {869, L42}

\bibitem[\protect\citeauthoryear{{Jim{\'e}nez} \& {Masset}}{{Jim{\'e}nez} \&
  {Masset}}{2017}]{Jimenez2017}
{Jim{\'e}nez} M.~A.,  {Masset} F.~S.,  2017, \mn@doi [\mnras]
  {10.1093/mnras/stx1946}, \href
  {https://ui.adsabs.harvard.edu/abs/2017MNRAS.471.4917J} {471, 4917}

\bibitem[\protect\citeauthoryear{{Keppler} et~al.,}{{Keppler}
  et~al.}{2018}]{Keppler2018}
{Keppler} M.,  et~al., 2018, \mn@doi [\aap] {10.1051/0004-6361/201832957},
  \href {https://ui.adsabs.harvard.edu/abs/2018A&A...617A..44K} {617, A44}

\bibitem[\protect\citeauthoryear{{Li}, {Colgate}, {Wendroff}  \& {Liska}}{{Li}
  et~al.}{2001}]{Li2001}
{Li} H.,  {Colgate} S.~A.,  {Wendroff} B.,   {Liska} R.,  2001, \mn@doi [\apj]
  {10.1086/320241}, \href
  {https://ui.adsabs.harvard.edu/#abs/2001ApJ...551..874L} {551, 874}

\bibitem[\protect\citeauthoryear{{Lin} \& {Papaloizou}}{{Lin} \&
  {Papaloizou}}{2010}]{Lin2010}
{Lin} M.-K.,  {Papaloizou} J. C.~B.,  2010, \mn@doi [\mnras]
  {10.1111/j.1365-2966.2010.16560.x}, \href
  {https://ui.adsabs.harvard.edu/abs/2010MNRAS.405.1473L} {405, 1473}

\bibitem[\protect\citeauthoryear{{Long} et~al.,}{{Long}
  et~al.}{2018}]{Long2018a}
{Long} F.,  et~al., 2018, \mn@doi [\apj] {10.3847/1538-4357/aae8e1}, \href
  {https://ui.adsabs.harvard.edu/abs/2018ApJ...869...17L} {869, 17}

\bibitem[\protect\citeauthoryear{{Lovelace} \& {Hohlfeld}}{{Lovelace} \&
  {Hohlfeld}}{2013}]{lovelace2013}
{Lovelace} R.~V.~E.,  {Hohlfeld} R.~G.,  2013, \mn@doi [\mnras]
  {10.1093/mnras/sts361}, \href
  {http://adsabs.harvard.edu/abs/2013MNRAS.429..529L} {429, 529}

\bibitem[\protect\citeauthoryear{{Lovelace}, {Li}, {Colgate}  \&
  {Nelson}}{{Lovelace} et~al.}{1999}]{Lovelace1999}
{Lovelace} R.~V.~E.,  {Li} H.,  {Colgate} S.~A.,   {Nelson} A.~F.,  1999,
  \mn@doi [\apj] {10.1086/306900}, \href
  {https://ui.adsabs.harvard.edu/#abs/1999ApJ...513..805L} {513, 805}

\bibitem[\protect\citeauthoryear{{Lyra}, {Turner}  \& {McNally}}{{Lyra}
  et~al.}{2015}]{Lyra15}
{Lyra} W.,  {Turner} N.~J.,   {McNally} C.~P.,  2015, \mn@doi [\aap]
  {10.1051/0004-6361/201424919}, \href
  {http://adsabs.harvard.edu/abs/2015A%26A...574A..10L} {574, A10}

\bibitem[\protect\citeauthoryear{{Masset}}{{Masset}}{2000}]{Masset2000}
{Masset} F.,  2000, \mn@doi [\aaps] {10.1051/aas:2000116}, \href
  {http://adsabs.harvard.edu/abs/2000A\%26AS..141..165M} {141, 165}

\bibitem[\protect\citeauthoryear{{Masset}}{{Masset}}{2002}]{Masset2002}
{Masset} F.~S.,  2002, \mn@doi [\aap] {10.1051/0004-6361:20020240}, \href
  {https://ui.adsabs.harvard.edu/abs/2002A&A...387..605M} {387, 605}

\bibitem[\protect\citeauthoryear{{Masset}}{{Masset}}{2008}]{Masset08}
{Masset} F.~S.,  2008, in {Goupil} M.~J.,  {Zahn} J.~P.,  eds,  EAS
  Publications Series Vol. 29, EAS Publications Series. pp 165--244,
  \mn@doi{10.1051/eas:0829006}

\bibitem[\protect\citeauthoryear{{Masset} \& {Papaloizou}}{{Masset} \&
  {Papaloizou}}{2003}]{Masset2003}
{Masset} F.~S.,  {Papaloizou} J.~C.~B.,  2003, \mn@doi [\apj] {10.1086/373892},
  \href {http://adsabs.harvard.edu/abs/2003ApJ...588..494M} {588, 494}

\bibitem[\protect\citeauthoryear{{Masset}, {D'Angelo}  \& {Kley}}{{Masset}
  et~al.}{2006}]{Masset2006}
{Masset} F.~S.,  {D'Angelo} G.,   {Kley} W.,  2006, \mn@doi [\apj]
  {10.1086/507515}, \href {http://adsabs.harvard.edu/abs/2006ApJ...652..730M}
  {652, 730}

\bibitem[\protect\citeauthoryear{{McNally}, {Nelson}, {Paardekooper}  \&
  {Ben{\'{\i}}tez-Llambay}}{{McNally} et~al.}{2019}]{McNally2019}
{McNally} C.~P.,  {Nelson} R.~P.,  {Paardekooper} S.-J.,
  {Ben{\'{\i}}tez-Llambay} P.,  2019, \mn@doi [\mnras] {10.1093/mnras/stz023},
  \href {http://adsabs.harvard.edu/abs/2019MNRAS.484..728M} {484, 728}

\bibitem[\protect\citeauthoryear{{Meru}, {Rosotti}, {Booth}, {Nazari}  \&
  {Clarke}}{{Meru} et~al.}{2019}]{Meru2019}
{Meru} F.,  {Rosotti} G.~P.,  {Booth} R.~A.,  {Nazari} P.,   {Clarke} C.~J.,
  2019, \mn@doi [\mnras] {10.1093/mnras/sty2847}, \href
  {http://cdsads.u-strasbg.fr/abs/2019MNRAS.482.3678M} {482, 3678}

\bibitem[\protect\citeauthoryear{{M{\"u}ller}, {Kley}  \& {Meru}}{{M{\"u}ller}
  et~al.}{2012}]{Muller2012}
{M{\"u}ller} T.~W.~A.,  {Kley} W.,   {Meru} F.,  2012, \mn@doi [\aap]
  {10.1051/0004-6361/201118737}, \href
  {https://ui.adsabs.harvard.edu/abs/2012A&A...541A.123M} {541, A123}

\bibitem[\protect\citeauthoryear{{Nazari}, {Booth}, {Clarke}, {Rosotti},
  {Tazzari}, {Juhasz}  \& {Meru}}{{Nazari} et~al.}{2019}]{Nazari2019}
{Nazari} P.,  {Booth} R.~A.,  {Clarke} C.~J.,  {Rosotti} G.~P.,  {Tazzari} M.,
  {Juhasz} A.,   {Meru} F.,  2019, \mn@doi [\mnras] {10.1093/mnras/stz836},
  \href {http://cdsads.u-strasbg.fr/abs/2019MNRAS.485.5914N} {485, 5914}

\bibitem[\protect\citeauthoryear{{Okuzumi}, {Momose}, {Sirono}, {Kobayashi}  \&
  {Tanaka}}{{Okuzumi} et~al.}{2016}]{Okuzumi2016}
{Okuzumi} S.,  {Momose} M.,  {Sirono} S.-i.,  {Kobayashi} H.,   {Tanaka} H.,
  2016, \mn@doi [\apj] {10.3847/0004-637X/821/2/82}, \href
  {https://ui.adsabs.harvard.edu/abs/2016ApJ...821...82O} {821, 82}

\bibitem[\protect\citeauthoryear{{Paardekooper}}{{Paardekooper}}{2014}]{Paardekooper2014}
{Paardekooper} S.~J.,  2014, \mn@doi [\mnras] {10.1093/mnras/stu1542}, \href
  {https://ui.adsabs.harvard.edu/abs/2014MNRAS.444.2031P} {444, 2031}

\bibitem[\protect\citeauthoryear{{Paardekooper}, {Baruteau}  \&
  {Kley}}{{Paardekooper} et~al.}{2011}]{Paardekooper2011}
{Paardekooper} S.~J.,  {Baruteau} C.,   {Kley} W.,  2011, \mn@doi [\mnras]
  {10.1111/j.1365-2966.2010.17442.x}, \href
  {https://ui.adsabs.harvard.edu/abs/2011MNRAS.410..293P} {410, 293}

\bibitem[\protect\citeauthoryear{{Pepli{\'n}ski}, {Artymowicz}  \&
  {Mellema}}{{Pepli{\'n}ski} et~al.}{2008a}]{Peplinski2008b}
{Pepli{\'n}ski} A.,  {Artymowicz} P.,   {Mellema} G.,  2008a, \mn@doi [\mnras]
  {10.1111/j.1365-2966.2008.13046.x}, \href
  {https://ui.adsabs.harvard.edu/abs/2008MNRAS.386..179P} {386, 179}

\bibitem[\protect\citeauthoryear{{Pepli{\'n}ski}, {Artymowicz}  \&
  {Mellema}}{{Pepli{\'n}ski} et~al.}{2008b}]{Peplinski2008c}
{Pepli{\'n}ski} A.,  {Artymowicz} P.,   {Mellema} G.,  2008b, \mn@doi [\mnras]
  {10.1111/j.1365-2966.2008.13339.x}, \href
  {https://ui.adsabs.harvard.edu/abs/2008MNRAS.387.1063P} {387, 1063}

\bibitem[\protect\citeauthoryear{{P{\'e}rez}, {Casassus}  \&
  {Ben{\'\i}tez-Llambay}}{{P{\'e}rez} et~al.}{2018}]{Perez2018}
{P{\'e}rez} S.,  {Casassus} S.,   {Ben{\'\i}tez-Llambay} P.,  2018, \mn@doi
  [\mnras] {10.1093/mnrasl/sly109}, \href
  {https://ui.adsabs.harvard.edu/abs/2018MNRAS.480L..12P} {480, L12}

\bibitem[\protect\citeauthoryear{{P{\'e}rez}, {Casassus}, {Baruteau}, {Dong},
  {Hales}  \& {Cieza}}{{P{\'e}rez} et~al.}{2019}]{Perez2019}
{P{\'e}rez} S.,  {Casassus} S.,  {Baruteau} C.,  {Dong} R.,  {Hales} A.,
  {Cieza} L.,  2019, \mn@doi [\aj] {10.3847/1538-3881/ab1f88}, \href
  {https://ui.adsabs.harvard.edu/abs/2019AJ....158...15P} {158, 15}

\bibitem[\protect\citeauthoryear{{Pierens} \& {Hur{\'e}}}{{Pierens} \&
  {Hur{\'e}}}{2005}]{PH05}
{Pierens} A.,  {Hur{\'e}} J.-M.,  2005, \mn@doi [\aap]
  {10.1051/0004-6361:200500099}, \href
  {https://ui.adsabs.harvard.edu/abs/2005A%26A...433L..37P} {433, L37}

\bibitem[\protect\citeauthoryear{{Pierens} \& {Lin}}{{Pierens} \&
  {Lin}}{2018}]{PierensLin2018}
{Pierens} A.,  {Lin} M.-K.,  2018, \mn@doi [\mnras] {10.1093/mnras/sty1314},
  \href {http://cdsads.u-strasbg.fr/abs/2018MNRAS.479.4878P} {479, 4878}

\bibitem[\protect\citeauthoryear{{Pierens} \& {Raymond}}{{Pierens} \&
  {Raymond}}{2016}]{PierensRaymond16}
{Pierens} A.,  {Raymond} S.~N.,  2016, \mn@doi [\mnras]
  {10.1093/mnras/stw1904}, \href
  {https://ui.adsabs.harvard.edu/abs/2016MNRAS.462.4130P} {462, 4130}

\bibitem[\protect\citeauthoryear{{Pinte}, {Dent}, {M{\'e}nard}, {Hales},
  {Hill}, {Cortes}  \& {de Gregorio-Monsalvo}}{{Pinte}
  et~al.}{2016}]{Pinte2016}
{Pinte} C.,  {Dent} W.~R.~F.,  {M{\'e}nard} F.,  {Hales} A.,  {Hill} T.,
  {Cortes} P.,   {de Gregorio-Monsalvo} I.,  2016, \mn@doi [\apj]
  {10.3847/0004-637X/816/1/25}, \href
  {https://ui.adsabs.harvard.edu/abs/2016ApJ...816...25P} {816, 25}

\bibitem[\protect\citeauthoryear{{Riols} \& {Lesur}}{{Riols} \&
  {Lesur}}{2019}]{RiolsLesur19}
{Riols} A.,  {Lesur} G.,  2019, \mn@doi [\aap] {10.1051/0004-6361/201834813},
  \href {https://ui.adsabs.harvard.edu/abs/2019A%26A...625A.108R} {625, A108}

\bibitem[\protect\citeauthoryear{{Shakura} \& {Sunyaev}}{{Shakura} \&
  {Sunyaev}}{1973}]{Shakura1973}
{Shakura} N.~I.,  {Sunyaev} R.~A.,  1973, \aap, \href
  {http://adsabs.harvard.edu/abs/1973A\%26A....24..337S} {24, 337}

\bibitem[\protect\citeauthoryear{{Simon}, {Lesur}, {Kunz}  \&
  {Armitage}}{{Simon} et~al.}{2015}]{Simon15}
{Simon} J.~B.,  {Lesur} G.,  {Kunz} M.~W.,   {Armitage} P.~J.,  2015, \mn@doi
  [\mnras] {10.1093/mnras/stv2070}, \href
  {http://cdsads.u-strasbg.fr/abs/2015MNRAS.454.1117S} {454, 1117}

\bibitem[\protect\citeauthoryear{{Teague}, {Bae}, {Bergin}, {Birnstiel}  \&
  {Foreman-Mackey}}{{Teague} et~al.}{2018}]{Teague2018}
{Teague} R.,  {Bae} J.,  {Bergin} E.~A.,  {Birnstiel} T.,   {Foreman-Mackey}
  D.,  2018, \mn@doi [\apjl] {10.3847/2041-8213/aac6d7}, \href
  {https://ui.adsabs.harvard.edu/abs/2018ApJ...860L..12T} {860, L12}

\bibitem[\protect\citeauthoryear{{Varni{\`e}re} \& {Tagger}}{{Varni{\`e}re} \&
  {Tagger}}{2006}]{Varniere2006}
{Varni{\`e}re} P.,  {Tagger} M.,  2006, \mn@doi [\aap]
  {10.1051/0004-6361:200500226}, \href
  {https://ui.adsabs.harvard.edu/abs/2006A&A...446L..13V} {446, L13}

\bibitem[\protect\citeauthoryear{{Weber}, {P{\'e}rez}, {Ben{\'\i}tez-Llambay},
  {Gressel}, {Casassus}  \& {Krapp}}{{Weber} et~al.}{2019}]{Weber2019}
{Weber} P.,  {P{\'e}rez} S.,  {Ben{\'\i}tez-Llambay} P.,  {Gressel} O.,
  {Casassus} S.,   {Krapp} L.,  2019, \mn@doi [\apj]
  {10.3847/1538-4357/ab412f}, \href
  {https://ui-adsabs-harvard-edu.insu.bib.cnrs.fr/abs/2019ApJ...884..178W}
  {884, 178}

\bibitem[\protect\citeauthoryear{{Zhang} et~al.,}{{Zhang}
  et~al.}{2018}]{Zhang2018}
{Zhang} S.,  et~al., 2018, \mn@doi [\apjl] {10.3847/2041-8213/aaf744}, \href
  {https://ui.adsabs.harvard.edu/abs/2018ApJ...869L..47Z} {869, L47}

\bibitem[\protect\citeauthoryear{{Zhu} \& {Baruteau}}{{Zhu} \&
  {Baruteau}}{2016}]{Zhu2016}
{Zhu} Z.,  {Baruteau} C.,  2016, \mn@doi [\mnras] {10.1093/mnras/stw202}, \href
  {http://adsabs.harvard.edu/abs/2016MNRAS.458.3918Z} {458, 3918}

\makeatother
\end{thebibliography}

\appendix
\section{Additional results for fast and slow migration}
\label{sec:appendix_a}
\begin{figure*}
\begin{center}
\includegraphics[width=0.99\hsize]{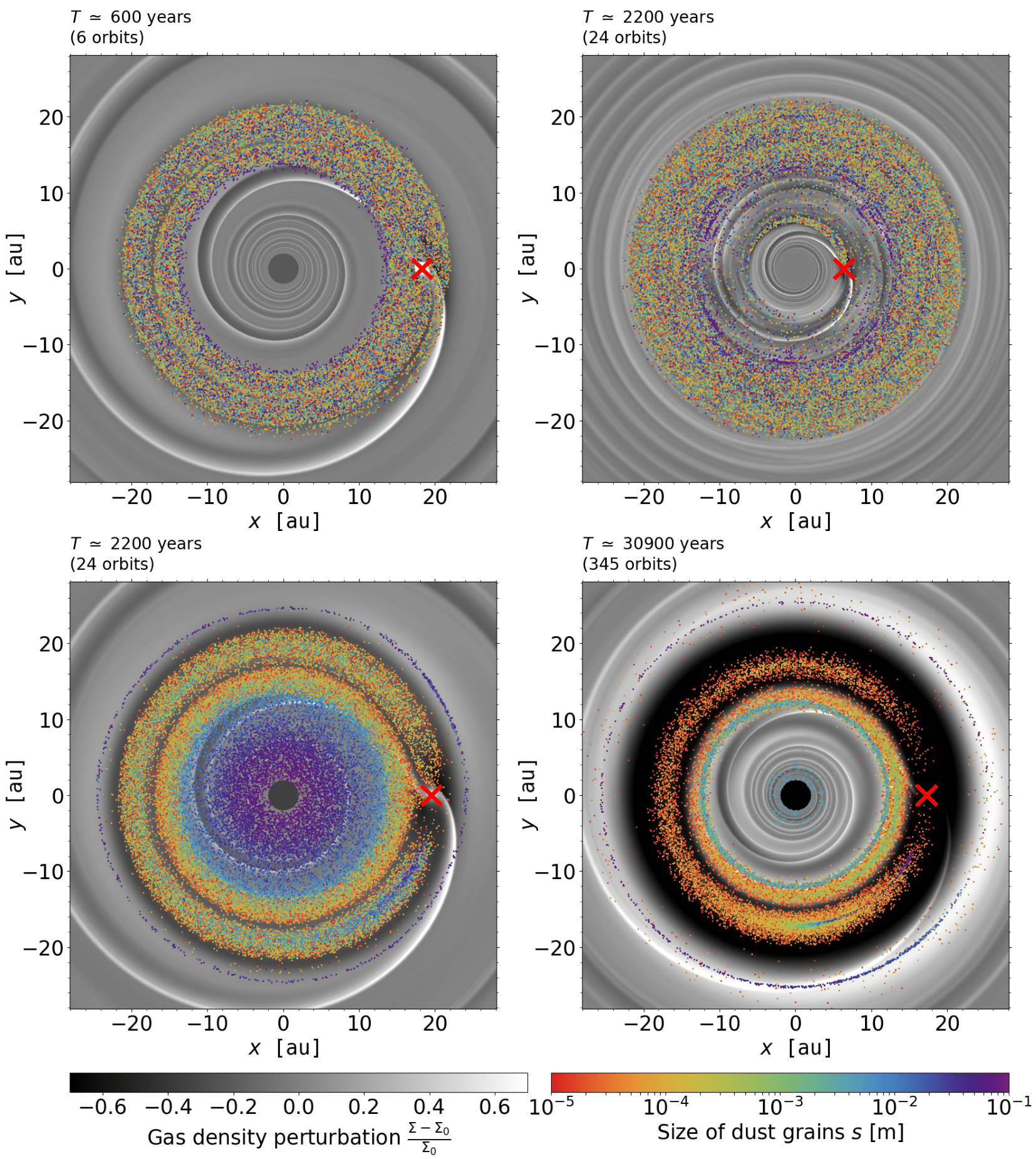}
\end{center}
\caption{Perturbed gas surface density relative to its initial profile, ($\Sigma-\Sigma_0)/\Sigma_0$, in black and white and in Cartesian coordinates. The location of the dust particles is overlaid by colored dots (see rainbow color bar below the bottom-right panel). In each panel, the red cross marks the position of the planet. Results are shown for a simulation with a disk setup similar to that of the Pendragon run, but for two different initial surface densities: $\Sigma_0=3\times10^{-3}$ (upper panels, fast planet migration) and $\Sigma_0=3\times10^{-4}$ (lower panels, slow migration).}
\label{fig:appendix}
\end{figure*}

We have seen in Section~\ref{sec:overview} via Fig.~\ref{fig:general} that for our disk model and range of planet masses, migration could be either fast, slow or intermittent, depending on the background surface density of the gas. While the main body of the paper focuses on intermittent migration, we illustrate in this section our main findings in the fast and slow migration regimes. Results are shown in Fig.~\ref{fig:appendix} for the same disk setup as in the Pendragon simulation (see Table~\ref{table:case2_parameters}), but for different initial surface densities of the gas at 10 au ($\Sigma_0$).

The upper panels are for $\Sigma_0=3\times10^{-3}$, that is for the fast migration case highlighted in Section~\ref{sec:overview}. After $\mathcal{P}$ has moved through the initial dust band (upper-left panel), some particles get trapped around the L4 Lagrange point located ahead of $\mathcal{P}$ in the azimuthal direction, which is clearly seen in the upper-right panel. Comparison between both panels shows that, apart from the dust trapped near L4, most of the dust particles ultimately forms a band behind $\mathcal{P}$ that is similar to the initial one, except that it drifts radially due to gas drag, and diffuses radially due to turbulence.

The lower panels are for $\Sigma_0=3\times10^{-4}$ (slow migration). The bottom-right panel shows that the planet forms three rings: (i) an outer ring that is mostly comprised of large particles, and which roughly stays at its initial location (compare with the bottom-left panel), (ii) a coorbital ring, and (iii) an inner ring that builds up at the pressure maximum between the main gap around the planet orbit and a secondary gap inside the orbit. Concerning the outer ring, a different choice of initial particles location could have led to more particles being trapped there, which would have made this ring more visible and/or thicker. Both the coorbital and the inner rings move inward together with the planet. Note that the inner ring only forms at low viscosities (it does not form for $\alpha=10^{-3}$). The bottom-right panel indicates that some large particles remain in the inner disk parts. We attribute this seeming trapping to a numerical artefact of the grid's inner wave-killing zone rather than to a physical pressure maximum arising from a tertiary gap carved by the planet's inner wake.

\section{A simple modification to the expression of the coorbital vortensity deficit for fast migration}
\label{sec:appendix_b}

The aim of this section is to propose a simple generalized expression for the coorbital vortensity deficit $\delta m$ when the assumption of slow migration in Eq.~(\ref{eq:slowspeed}) is not satisfied. We shall still assume, however, that $\delta m$ can be defined by Eq.~(\ref{eq:deficit}). By dividing the right-hand side of Eq.~(\ref{eq:deficit}) by $2\pi r_{\rm p}$, and integrating over $\theta$, we get
\begin{equation}
\begin{aligned}
\begin{split}
\delta m \approx 4\pi r_{\rm p}\omega(r_{\rm p}) & \times
\left[\frac{x_{\rm s}}{2\pi r_{\rm p}}\int_{\theta_1}^{\theta_2}\mathcal{I_V}\left(r_{\rm p}-x_{\rm s},\theta\right) r_{\rm p} d\theta \right. \\
& - \left. \frac{1}{2\pi r_{\rm p}} \int_{\theta_1}^{\theta_2} \int_{r_{\rm p}-x_{\rm s}}^{r_{\rm p}} \mathcal{I_V}\left(r,\theta\right) r_{\rm p} d\theta dr\right],
\end{split}
\end{aligned}
\end{equation}
which assumes $r \approx r_{\rm p}$, and where [$\theta_1,\theta_2$] defines the azimuthal extent of the material trapped in libration with $\mathcal{P}$ and of the orbit-crossing flow ($\Delta \theta = \theta_2-\theta_1$). Further assuming that the material trapped in libration has uniform $\mathcal{I_V}$ over a rectangular domain of extent $2 x_{\rm s} \times \Delta \theta$ in polar cylindrical coordinates, and that $\mathcal{I_V}\left(r_{\rm p}-x_{\rm s},\theta\right)$ is independent of $\theta$ over [$\theta_1,\theta_2$] (see Fig.~\ref{fig:invortensity} and lower-left panel in Fig.~\ref{fig:psig0}), Eq.~(\ref{eq:deficit}) can be recast as
\begin{equation}
\begin{aligned}
\delta m & \approx 4\pi r_{\rm p}\omega(r_{\rm p}) \times
\left[\frac{x_{\rm s}}{2\pi r_{\rm p}}\mathcal{I_V}_e r_{\rm p} \Delta \theta - \frac{1}{2\pi r_{\rm p}}
\mathcal{I_V}_{\rm lib} r_{\rm p} \Delta \theta x_{\rm s}\right] \\
& \approx 4\pi r_{\rm p}\omega(r_{\rm p})x_{\rm s} \times
\frac{\Delta \theta}{2\pi} \times
\left[\mathcal{I_V}_{e}-
\mathcal{I_V}_{\rm lib}\right].
\label{eq:appendix_deficit}
\end{aligned}
\end{equation}
Recall that $\mathcal{I_V}_{e}$ is the inverse vortensity of the gas entering the horseshoe region ($\mathcal{I_V}_{e} = \mathcal{I_V}(r_{\rm p}-x_{\rm s})$), and $\mathcal{I_V}_{\rm lib}$ that of the gas trapped in libration with $\mathcal{P}$. The r.h.s. of Eq.~(\ref{eq:appendix_deficit}) is therefore that of Eq.~(\ref{eq:deficit_approx}) multiplied by $\Delta \theta/2\pi$.

\bsp	
\label{lastpage}
\end{document}